\documentclass[aps,preprint,psfig,epsf]{revtex4}
\usepackage{graphicx}
\usepackage{amssymb}
\usepackage{amsmath}

\input{tcilatex}

\begin{document}

\title{The two-loop functional renormalization-group approach to the one-
and two-dimensional Hubbard model}
\author{A. A. Katanin$^{a,b}$}
\affiliation{$^{a}$Max-Planck Institute f\"ur Festk\"orperforschung, 70569, Stuttgart,
Germany\\
$^{b}$Institute of Metal Physics, 620044, Ekaterinburg, Russia}

\begin{abstract}
We consider the application of the two-loop functional renormalization-group
(fRG) approach to study the low-dimensional Hubbard model. This approach
accounts for both, the universal and non-universal contributions to the RG
flow. While the universal contributions were studied previously within the
field-theoretical RG for the one-dimensional Hubbard model with linearized
electronic dispersion and the two-dimensional Hubbard model with flat Fermi
surface, the non-universal contributions to the flow of the vertices and
susceptibilities appear to be important at large momenta scales. The
two-loop fRG approach is also applied to the two-dimensional Hubbard model
with a curved Fermi surface and the van Hove singularities near the Fermi
level. The vertices and susceptibilities in the end of the flow of the
two-loop approch are suppressed in comparison with both the one-loop
approach with vertex projection and the modified one-loop approach with
corrected vertex projection errors. The results of the two-loop approach are
closer to the results of one-loop approach with the projected vertices, the
difference of the results of one- and two-loop fRG approaches decreases when
going away from the van Hove band filling. The quasiparticle weight remains
finite in two dimensions at not too low temperatures above the paramagnetic
ground state.
\end{abstract}

\maketitle

\address{$^a$ Max-Planck-Institut f\"ur Festk\"orperforschung, D-70569,\\
Stuttgart\\
$^b$ Institute of Metal Physics, 620219 Ekaterinburg, Russia}

\section{Introduction}

The discovery of high-$T_{c}$ superconductors, which demonstrate nontrivial
properties in a broad temperature- and concentration range has dramatically
increased interest to correlated low-dimensional systems, and investigation
of these systems became a challenge for modern solid state physics. Later
discovery of unconventional triplet superconductors (in particular Sr$_{2}$%
RuO$_{4}$) has attracted further attention to a possible instabilities of a
Fermi-liquid state in the low-dimensional systems due to electronic
correlations. These compounds stimulate theoretical interest to study the
effect of correlations on electronic and magnetic properties of
low-dimensional systems.

The common model to treat electronic correlations is the one-band Hubbard
model. At sufficiently large on-site Coulomb repulsion $U\sim W$ ($W$ is the
bandwidth) this model describes the Mott metal-insulator transition. This
transition is an essentially non-perturbative phenomenon and is well
described by the dynamical mean-field theory \cite{DMFT}, which considers
the limit of infinite number of dimensions and neglects spatial
correlations. However, even in the weak- and intermediate coupling regime $%
U<W$ the Hubbard model is nontrivial in two dimensions near some special
(van Hove) band fillings or Fermi surface nesting, where magnetic and/or
superconducting instabilities may arise \cite{LinHirsch}. The spatial
correlations, not accounted in the dynamical mean field theory, become
important in the vicinity of the corresponding quantum phase transitions.
Therefore, the development of methods, which are able to describe magnetic
or superconducting fluctuations is of high interest.

While the 1D Hubbard model is exactly solvable by the Bethe ansatz and the
phase diagram of this model with linearized electronic dispersion was
obtained within the bosonization and field-theoretical renormalization group
methods \cite{Solyom}; other numerical or approximate methods have to be
used in higher dimensions. The applicability of numerical methods (exact
diagonalization, quantum Monte Carlo (QMC), dynamical cluster approximations
etc.) which treat spatial correlations is restricted by the cluster size
and/or not too low temperatures. At the same time, there is a number of
different (semi-)analytical approximations which treat the Hubbard model in
the weak- and intermediate-coupling regime. The simplest is the mean-field
approximation which treats the electron-electron interaction via some
effective field applied to the fermionic system \cite{LinHirsch}. Regarding
the stability of the paramagnetic state this approach is essentially
equivalent to the requirement that the corresponding susceptibilities in the
random phase approximation (RPA) or $T$-matrix approximation (TMA) \cite%
{Kanamori} remain positive and finite in the absence of instabilities in the
particle-hole (ph) or particle-particle (pp) channel. The corresponding
electron-electron interaction vertex, irreducible in the ph or the pp
channel is supposed to be equal to the bare on-site Coulomb repulsion $U$ in
these approaches$.$

More complicated approaches account for the effect of fluctuations. These
approaches can be subdivided into two classes: (i) approaches which consider
the effect of the renormalization of the ph- or pp-irreducible
electron-electron interaction vertex and (ii) approaches which consider in
addition to (i) the renormalization of the one-particle Green functions. One
of the approximations of the first class is the combination of RPA and TMA,
which was proposed to account for both, the ph- and the pp-scattering \cite%
{Libsch,KT}. In particular, one can use the RPA vertex (instead of the bare $%
U$) as the pp-irreducible vertex in TMA \cite{Libsch}, or, vice versa, one
can use the TMA vertex, which is irreducible in particle-hole channel,
instead of the bare vertex in RPA \cite{KT}. The two-particle
self-consistent (TPSC) approximation \cite{TPSC} uses the RPA-type vertex
with the effective interaction $U_{ef}$ instead of the bare one, the $U_{ef}$
is determined by the requirement of the fulfillment of sum rules.

On the other hand, the commonly used approximation of the class (ii) is the
FLEX approximation \cite{FLEX}. This approximaton uses the RPA interaction
vertices but accounts for the renormalization of the one-particle Green
functions as well. More complex approach of the type (ii) is the parquet
approach \cite{Bickers1,JanisR,Janis1,DzyJak} which considers the
contribution of different channels of electron scattering and their mutual
interplay in the interaction vertex. However, the practical application of
this approach for systems with the dimensionality $d>1$ meets serious
computational difficulties and was performed only in few cases \cite%
{Janis1,DzyJak}. The abovementioned approximations give a possibility to
treat spatial correlations of the Hubbard model in the weak- and
intermediate coupling regime. However, the accuracy of the results obtained
within these approximations can be hardly controlled; 
these approximations can also be hardly improved.

The recently proposed functional renormalization-group (fRG) approaches \cite%
{Polchinskii,Zanchi,Metzner,SalmBook,SalmHon,SalmHonR,SalmHon1} use a
different strategy. Integrating out modes with quasiparticle (qp) energy $%
|\varepsilon _{\mathbf{k}}|\geq \Lambda ,$ where $\Lambda $ is the cutoff
parameter, one obtains a (formally exact) hierarchy of RG equations for the $%
n$-particle interaction vertices. This hierarchy is usually truncated by
neglecting higher-order vertices. To leading (one-loop) order these
equations neglect the six-point vertex and describe the renormalization of
the two-particle electron-electron interaction vertices only. Therefore the
one-loop fRG approach belongs to approximations of class (i). Unlike the RPA
and TMA, however, different electron scattering channels are treated on the
same footing within the fRG. In one dimension this approach allows to
reproduce the results obtained earliar within the field-theoretical RG
approach \cite{IntFlow}. The results for the instabilities, flow of
electron-electron interaction vertices, and phase diagrams of the standard 
\cite{Zanchi,Metzner,SalmHon,SalmHon1,IntFlow,Our2D} as well as the extended 
\cite{Our2DUVJ} 2D Hubbard model were also obtained at one-loop order.

The self-energy effects, which are not included in the one-loop
calculations, can be consistently taken into account at the two-loop order.
In one dimension these effects are shown to be crucially important to
describe Luttinger liquid behavior\cite{Solyom}. The calculation of the
scattering rates \cite{SalmHon}, quasiparticle residues \cite{SalmHonSE} and
the electronic self-energy \cite{KK,Metz} in two dimensions using vertices
obtained in the one-loop approximation showed, however, that contrary to the
1D case the self-energy effects in 2D are much weaker.

To estimate corrections to the one-loop approximation, however, the full
calculation of the two-loop contributions to the flow of vertices is
necessary. Contrary to the calculations at one-loop order, the two-loop
corrections account partly for the frequency dependence of the vertices and
their momentum dependence beyond the projection to the Fermi surface.
Therefore, the two-loop calculations serve also as a test of the importance
of the frequency- and momentum dependence of the vertices. Finally, they
provide an information about quasiparticle weight, damping and the
interaction-induced Fermi surface shifts.

Although the two-loop corrections were considered previously for 2D systems
in Ref. \cite{Freire} within the field-theoretical renormalization group
approach, the application of this approach is limited to flat Fermi surfaces
and the electronic dispersion linearized near the Fermi surface. The
advantage of the functional renormalization group approach is that it can be
applied to both, flat and curved Fermi surfaces with or without van Hove
singularities, since this method does not require universality of the
scaling functions. The applicability of this approach for calculation of the
two-loop corrections to scaling functions of the bosonic $\phi ^{4}$ model
was investigated in Ref. \cite{Kopietz} where the need of account of the
momentum- and frequency dependence of the vertices was emphasized. The
treatment of this dependence numerically is, however, a rather difficult
task.

In the present paper we use a slightly different method, which allows us to
avoid considering momentum and frequency dependence of the higher-order
vertices and calculate the two-loop corrections and investigate their
influence on the flow of the coupling constants, susceptibilities and
self-energies of the Hubbard model. We use the momentum-cutoff version of
the fRG for the 1PI functions, which is applicable in the vicinity of
antiferromagnetic or superconducting phase$.$

The plan of the paper is the following. In Sect. 2 we introduce and compare
the one- and two-loop fRG approaches. In Sect. 3 we apply the two-loop fRG
approach to the 1D and 2D Hubbard models and investigate the flow of the
interaction vertices and susceptibilities in this approach. In conclusion
(Sect. 4), we discuss results of the paper and outline future perspectives
of the method. The derivation of the two-loop equations is presented in the
Appendix.

\section{The model and the two-loop fRG approach}

We consider the Hubbard model%
\begin{equation}
H=-\sum_{ij\sigma }t_{ij}c_{i\sigma }^{\dagger }c_{j\sigma
}+U\sum_{i}n_{i\uparrow }n_{i\downarrow }-\mu n,  \label{H}
\end{equation}%
where the hopping amplitude $t_{ij}=t$ for nearest neighbor sites $i$ and $j$
and $t_{ij}=-t^{\prime }$ for next-nearest neighbor sites ($t,t^{\prime }>0$%
), $\mu $ is the chemical potantial, corresponding to the particle number $n$%
. In momentum space Eq. (\ref{H}) reads 
\begin{equation}
H=\sum_{\mathbf{k}\sigma }\varepsilon _{\mathbf{k}}c_{\mathbf{k}\sigma
}^{\dagger }c_{\mathbf{k}\sigma }+\frac{U}{2N}\sum_{\mathbf{k}_{1}\mathbf{k}%
_{2}\mathbf{k}_{3}\mathbf{k}_{4}}\sum_{\sigma \neq \sigma ^{\prime }}c_{%
\mathbf{k}_{1}\sigma }^{\dagger }c_{\mathbf{k}_{2}\sigma ^{\prime
}}^{\dagger }c_{\mathbf{k}_{3}\sigma ^{\prime }}c_{\mathbf{k}_{4}\sigma
}\delta _{\mathbf{k}_{1}+\mathbf{k}_{2}-\mathbf{k}_{3}-\mathbf{k}_{4}},
\label{H1}
\end{equation}%
where the Kronecker $\delta $-symbol ensures momentum conservation, $%
\varepsilon _{\mathbf{k}}$ is the electronic dispersion, $N$ is the number
of sites.

To calculate physical properties of the model (\ref{H}) we apply the fRG
approach with a sharp momentum cutoff (see, e.g. Ref. \cite{SalmHonSE}),
which considers an effective action obtained by integrating out modes with
the quasiparticle energy $|\varepsilon _{\mathbf{k}}|\geq \Lambda $, $%
\Lambda $ being the cutoff parameter. This procedure is especially
convenient when there is no ferromagnetic instability developing in the
weak-coupling regime (in two dimensions this implies $t^{\prime }\lesssim
0.3t$, see Refs. \cite{IKK,SalmHon1}). In case of a ferromagnetic
instability the contribution of small-momenta particle-hole scattering which
is not included in the momentum cutoff fRG approaches may become important
already at sufficiently large momenta scales \cite{IKK,SalmHon1}, this case
is not considered in the present paper. Among different versions of the fRG
approach (Polchinskii \cite{Polchinskii,Zanchi}, Wick-ordered \cite%
{SalmBook,Metzner} or one-particle irreducible (1PI) \cite%
{SalmHon,SalmHonR,SalmHonSE}), we use the fRG approach for the 1PI
functions. In this approach the electron propagator at scale $\Lambda $ has
the form%
\begin{equation}
G_{\Lambda }(\mathbf{k},i\nu _{n})=\frac{\theta (|\varepsilon _{\mathbf{k}%
}|-\Lambda )}{i\nu _{n}-\varepsilon _{\mathbf{k}}-\theta (|\varepsilon _{%
\mathbf{k}}|-\Lambda )\Sigma _{\Lambda }(\mathbf{k},i\nu _{n})}
\end{equation}%
where $\Sigma _{\Lambda }(\mathbf{k},i\nu _{n})$ is the self-energy at the
same scale, $\nu _{n}$ are the fermionic Matsubara frequencies$.$ For $%
\Lambda \geq \Lambda _{0}=\max (|\varepsilon _{\mathbf{k}}|)$ the internal
one-particle Green functions in all the diagrams are zero, so that the
renormalization of the physical quantities is absent: the effective
interaction $V_{\Lambda }$ coincides with the bare one and $\Sigma _{\Lambda
}(\mathbf{k},i\nu _{n})=0$. The self-energy $\Sigma _{\Lambda }(\mathbf{k}%
,i\omega _{n})$ as well as the electron-electron interaction vertex $%
V_{\Lambda }(k_{1},k_{2};k_{3},k_{4})$ ($k_{1},k_{2}$ and $k_{3},k_{4}$ are
the momenta- and frequencies of the incoming and outgoing electrons, $k_{i}=(%
\mathbf{k}_{i},i\nu _{n}^{(i)})$) at $\Lambda <\Lambda _{0}$ can be obtained
by integration of the corresponding flow equations.

At one-loop order 5 diagrams contribute to the renormalization of the
electron-electron interaction vertex $V_{\Lambda }(k_{1},k_{2};k_{3},k_{4})$
and two diagrams to the self-energy $\Sigma _{\Lambda }(\mathbf{k},i\nu
_{n}) $ (see Fig. 1). The corresponding flow equations can be written in the
form (see Refs. \cite{SalmHon,SalmHonR}) 
\begin{subequations}
\label{OneLoop}
\begin{eqnarray}
\frac{d\Sigma _{\Lambda }}{d\Lambda } &=&V_{\Lambda }\circ S_{\Lambda }
\label{OneLoopA} \\
\frac{dV_{\Lambda }}{d\Lambda } &=&V_{\Lambda }\circ (G_{\Lambda }\circ
S_{\Lambda }+S_{\Lambda }\circ G_{\Lambda })\circ V_{\Lambda }
\label{OneLoopB}
\end{eqnarray}%
\ where $\circ $ is the short notation for the summation over momentum-,
frequency- and spin-variables according to standard diagrammatic rules, see
diagrams of Fig. 1, the single-scale propagator $S_{\Lambda }(\mathbf{k}%
,i\nu _{n})$ is defined by 
\end{subequations}
\begin{equation}
S_{\Lambda }(\mathbf{k},i\nu _{n})=-\frac{\delta (|\varepsilon _{\mathbf{k}%
}|-\Lambda )}{i\nu _{n}-\varepsilon _{\mathbf{k}}-\theta (|\varepsilon _{%
\mathbf{k}}|-\Lambda )\Sigma _{\Lambda }(\mathbf{k},i\nu _{n})}.
\end{equation}%
The equations (\ref{OneLoop})\ should be solved with the initial conditions $%
V_{\Lambda _{0}}=U$ and $\Sigma _{\Lambda _{0}}=0$. To demonstrate how the
fRG equations (\ref{OneLoop}) reproduce the perturbation theory results, it
is helpful to expand their solution in the bare interaction $U$. To this
end, we solve them iteratively. Starting from the bare values of $V$ and $%
\Sigma $ we obtain after one iteration the first-order result for the
self-energy and the second-order perturbation theory (SOPT) result for the
vertex 
\begin{eqnarray}
\Sigma _{\Lambda }^{(1)} &=&U\text{Tr}(G_{\Lambda }^{0})  \label{SE1} \\
V_{\Lambda }^{(1)} &=&U+U^{2}(G_{\Lambda }^{0}\circ G_{\Lambda }^{0})  \notag
\end{eqnarray}%
where the index \textquotedblleft 0\textquotedblright\ stands for the bare
Green functions with $\Sigma =0.$ After the second iteration we have 
\begin{eqnarray}
\Sigma _{\Lambda }^{(2),1\text{-loop}} &=&U\text{Tr}(G_{\Lambda
}^{(1)})+U^{2}G_{\Lambda }^{0}\circ G_{\Lambda }^{0}\circ G_{\Lambda }^{0} 
\notag \\
V_{\Lambda }^{(2),1\text{-loop}} &=&U+U^{2}\int_{\Lambda }^{\Lambda
_{0}}d\Lambda ^{\prime }[S_{\Lambda ^{\prime }}^{(1)}\circ G_{\Lambda
^{\prime }}^{(1)}+G_{\Lambda ^{\prime }}^{(1)}\circ S_{\Lambda ^{\prime
}}^{(1)}]+U^{3}[G_{\Lambda }^{0}\circ G_{\Lambda }^{0}\circ G_{\Lambda
}^{0}\circ G_{\Lambda }^{0}]_{\text{ladder}}  \label{OneLoop1A} \\
&&+U^{3}\int_{\Lambda }^{\Lambda _{0}}d\Lambda ^{\prime }[(G_{\Lambda
^{\prime }}^{0}\circ G_{\Lambda ^{\prime }}^{0})_{\text{in}}\circ \frac{d}{%
d\Lambda ^{\prime }}(G_{\Lambda ^{\prime }}^{0}\circ G_{\Lambda ^{\prime
}}^{0})_{\text{ex}}]_{\text{non-ladder}}  \notag
\end{eqnarray}%
Here the $G^{(1)}$ and $S^{(1)}$ functions are calculated with the
self-energy (\ref{SE1}), \textquotedblleft ladder\textquotedblright\ and
\textquotedblleft non-ladder\textquotedblright\ denote two different kinds
of diagrams (see Fig. 2), \textquotedblleft in\textquotedblright\ and
\textquotedblleft ex\textquotedblright\ denote the Green functions which
belong to the internal and external bubble in the non-ladder diagrams, as
shown in Fig. 2. The integrands in the second and last line do not form a
total $\Lambda $-derivative, and therefore we do not obtain the exact
third-order perturbation theory (TOPT) result for the vertex. While the
integrand in the first line of the Eq. (\ref{OneLoop1A}) can be changed to
form the total derivative by the replacement $S_{\Lambda }\rightarrow
dG_{\Lambda }/d\Lambda $, which was was shown in Ref. \cite{Ward} to be
equivalent to borrowing some terms from the two-loop corrections to the
vertex, casting the term in the third line of Eq. (\ref{OneLoop1A}) to the
form of the total derivative requires full consideration of the two-loop
corrections.

The consideration above provides a definition of the $n$-loop\textit{\ }%
approximation\textit{\ }as an approximation\textit{\ }which\textit{\
correctly reproduces }$n$\textit{-loop parts of the diagrams for the
two-particle interaction vertex and }$n+1$\textit{-loop parts of the
self-energy diagrams}. In the presence of logarithic divergencies (e.g. in
one dimension), when $G_{\Lambda }^{0}\circ G_{\Lambda }^{0}\sim \ln
(\Lambda /\alpha )$ ($\alpha \ll \Lambda $ is a small parameter), the terms $%
(G_{\Lambda ^{\prime }}^{0}\circ G_{\Lambda ^{\prime }}^{0}\circ G_{\Lambda
^{\prime }}^{0})\circ (dG_{\Lambda ^{\prime }}^{0}/d\Lambda ^{\prime })$ and 
$U^{3}(G_{\Lambda ^{\prime }}^{0}\circ G_{\Lambda ^{\prime }}^{0})_{\text{ex}%
}\circ d(G_{\Lambda ^{\prime }}^{0}\circ G_{\Lambda ^{\prime }}^{0})_{\text{%
in}}/d\Lambda ^{\prime }$ which appear at the two-loop order (see below) and
are necessary to combine to the total derivatives in the Eq. (\ref{OneLoop1A}%
) can be neglected to leading logarithmic order. This provides another, more
conventional definition of the $n$-loop approximation as an approximation
which correctly treats the terms $U^{m}\ln ^{m-n}(\Lambda /\alpha )$ in the
perturbation series for the vertex ($m\geq n$). Note, however, that in two
dimensions the bubbles $G_{\Lambda ^{\prime }}^{0}\circ G_{\Lambda ^{\prime
}}^{0}$ are either non-divergent for arbitrary fillings or may contain
squared logarithmical divergencies for some special (van Hove) band
fillings, and the latter definition of the $n$-loop approximation can not be
applied.

To go beyond the one-loop order of the Eqs. (\ref{OneLoop}) one has to take
into account the contribution of the three-particle interaction vertex (see
Refs. \cite{SalmHon,SalmHonR} and Appendix). Generally, this vertex
generates contributions to the two-particle interaction vertex $V$ with an
arbitrary number of loops $n\geq 3.$ At this stage two different
approximations are possible. (a) keeping only contributions which are
necessary to treat exactly diagrams with fixed number $n$ of loops and (b)
keep all the contributions which are generated by an integration of the
equation for the $n+1$-particle vertex, neglecting $n+2$-particle vertex. In
the present paper we restrict ourselves to the approximation (a), i.e.
consider only those contributions to RG flow which are necessary to treat
exactly the two-loop parts of the diagrams.

At the two-loop order the flow equation for the self-energy (\ref{OneLoopA})
does not change while 32 new diagrams contribute to the renormalization of
the vertex (see Fig. 3). The resulting two-loop equations have the form (see
Appendix for the derivation) 
\begin{subequations}
\label{TwoLoop12}
\begin{eqnarray}
\frac{d\Sigma _{\Lambda }}{d\Lambda } &=&V_{\Lambda }\circ S_{\Lambda }
\label{TwoLoop1} \\
\frac{dV_{\Lambda }}{d\Lambda } &=&V_{\Lambda }\circ (G_{\Lambda }\circ
S_{\Lambda }+S_{\Lambda }\circ G_{\Lambda })\circ V_{\Lambda }
\label{TwoLoop2} \\
&&+S_{\Lambda }\circ \int\limits_{\Lambda }^{\Lambda _{0}}d\Lambda ^{\prime
}V_{\Lambda ^{\prime }}\circ G_{\Lambda ^{\prime }}\circ V_{\Lambda ^{\prime
}}\circ G_{\Lambda ^{\prime }}\circ V_{\Lambda ^{\prime }}\circ S_{\Lambda
^{\prime }}  \notag
\end{eqnarray}%
Contrary to the one-loop approximation the frequency dependence of the
vertex becomes essential at the two-loop order. This can be seen from the
fact that to reproduce the TOPT result one needs to iterate Eq. (\ref%
{TwoLoop2}) twice. If one neglects the frequency dependence of the vertices,
the equations (\ref{TwoLoop12}) fail to reproduce correct TOPT results. The
necessity of the account of the frequency and momentum dependence of the
vertex was previously emphasized in the two-loop calculation of the $\beta $%
-function of $\phi ^{4}$ theory \cite{Kopietz} and the self-energy
calculation in the 2D Hubbard model \cite{SalmHon2,SalmHonSE,KK,Metz}.

To avoid having explicitly the frequency- and momenta- dependent vertices,
we integrate Eq. (\ref{TwoLoop2}) formally and keep frequency- and momentum
dependence coming from the one-loop term only to obtain 
\end{subequations}
\begin{eqnarray}
V_{\Lambda } &=&\overline{V}_{\Lambda }+\delta V_{\Lambda }  \notag \\
\delta V_{\Lambda } &=&\int\limits_{\Lambda }^{\Lambda _{0}}d\Lambda
^{\prime }[V_{\Lambda ^{\prime }}\circ G_{\Lambda ^{\prime }}\circ
S_{\Lambda ^{\prime }}\circ V_{\Lambda ^{\prime }}-\widehat{\mathcal{P}}%
(V_{\Lambda ^{\prime }}\circ G_{\Lambda ^{\prime }}\circ S_{\Lambda ^{\prime
}}\circ V_{\Lambda ^{\prime }})]  \label{Sub}
\end{eqnarray}%
where $\overline{V}_{\Lambda }=\widehat{\mathcal{P}}V_{\Lambda }$ and the
operator $\widehat{\mathcal{P}}$ projects the external frequencies to zero
and the external momenta to the Fermi surface. Reinserting this vertex into
the one-loop contributions of Eqs. (\ref{TwoLoop12}) and using projected
vertices in the two-loop contributions, we obtain to linear order in $\delta
V$%
\begin{subequations}
\label{TwoLoopA}
\begin{eqnarray}
\frac{d\Sigma _{\Lambda }}{d\Lambda } &=&\overline{V}_{\Lambda }\circ
S_{\Lambda }+S_{\Lambda }\circ \int\limits_{\Lambda }^{\Lambda _{0}}d\Lambda
^{\prime }[\overline{V}_{\Lambda ^{\prime }}\circ G_{\Lambda ^{\prime
}}\circ S_{\Lambda ^{\prime }}\circ \overline{V}_{\Lambda ^{\prime }}-%
\widehat{\mathcal{P}}(\overline{V}_{\Lambda ^{\prime }}\circ G_{\Lambda
^{\prime }}\circ S_{\Lambda ^{\prime }}\circ \overline{V}_{\Lambda ^{\prime
}})]  \label{TwoLoopA1} \\
\frac{d\overline{V}_{\Lambda }}{d\Lambda } &=&\widehat{\mathcal{P}}\left\{ 
\overline{V}_{\Lambda }\circ (G_{\Lambda }\circ S_{\Lambda })\circ \overline{%
V}_{\Lambda }-\overline{V}_{\Lambda }\circ (G_{\Lambda }\circ S_{\Lambda
})\circ (\overline{V}_{\Lambda }^{1L}-V_{0})\right.  \notag \\
&&\ +\overline{V}_{\Lambda }\circ (G_{\Lambda }\circ S_{\Lambda })\circ
\int\limits_{\Lambda }^{\Lambda _{0}}d\Lambda ^{\prime }\overline{V}%
_{\Lambda ^{\prime }}\circ G_{\Lambda ^{\prime }}\circ S_{\Lambda ^{\prime
}}\circ \overline{V}_{\Lambda ^{\prime }}  \notag \\
&&\left. +S_{\Lambda }\circ \int\limits_{\Lambda }^{\Lambda _{0}}d\Lambda
^{\prime }\overline{V}_{\Lambda ^{\prime }}\circ G_{\Lambda ^{\prime }}\circ 
\overline{V}_{\Lambda ^{\prime }}\circ G_{\Lambda ^{\prime }}\circ \overline{%
V}_{\Lambda ^{\prime }}\circ S_{\Lambda ^{\prime }}\right\}
\label{TwoLoopA2}
\end{eqnarray}%
where 
\end{subequations}
\begin{equation*}
\overline{V}_{\Lambda }^{1L}=V_{0}+\widehat{\mathcal{P}}\int\limits_{\Lambda
}^{\Lambda _{0}}d\Lambda ^{\prime }(\overline{V}_{\Lambda ^{\prime }}\circ
G_{\Lambda ^{\prime }}\circ S_{\Lambda ^{\prime }}\circ \overline{V}%
_{\Lambda ^{\prime }})
\end{equation*}%
is the analogue of the 1-loop vertex calculated with the two-loop vertices $%
\overline{V}_{\Lambda ^{\prime }}.$ After this reinsertion, only the
projected vertices $\overline{V}$ enter the Eqs. (\ref{TwoLoopA}). While the
last term in the Eq. (\ref{TwoLoopA2}) accounts for the two-loop corrections
to the flow, other integral contributions correct the effect of the vertex
projection in one-loop diagrams. In particular, for $\Lambda $-independent $%
V $ and $\Sigma $ the last two terms in the Eq. (\ref{TwoLoopA2}) combine to
a $\Lambda $-derivative of the corresponding two-loop diagram, so that these
equations with the initial condition $\overline{V}_{\Lambda _{0}}^{1L}=%
\overline{V}_{\Lambda _{0}}=V_{0}\equiv U,$ $\Sigma _{\Lambda _{0}}=0$
correctly reproduce the result of the TOPT after one iteration. The two-loop
fRG equation for the self-energy (\ref{TwoLoopA1}) is identical to that
investigated earliar with one-loop vertices \cite{KK}. The flow of the
susceptibilities is described by the equation, similar to the Eq. (\ref%
{TwoLoopA2}), 
\begin{subequations}
\label{TwoLoopHi}
\begin{eqnarray}
\frac{d\chi _{\Lambda }}{d\Lambda } &=&T_{\Lambda }\circ (G_{\Lambda }\circ
S_{\Lambda })\circ T_{\Lambda } \\
\frac{dT_{\Lambda }}{d\Lambda } &=&\widehat{\mathcal{P}}\left\{ T_{\Lambda
}\circ (G_{\Lambda }\circ S_{\Lambda })\circ \overline{V}_{\Lambda
}-T_{\Lambda }\circ (G_{\Lambda }\circ S_{\Lambda })\circ (\overline{V}%
_{\Lambda }^{1L}-V_{0})\right.  \notag \\
&&\ +T_{\Lambda }\circ (G_{\Lambda }\circ S_{\Lambda })\circ
\int\limits_{\Lambda }^{\Lambda _{0}}d\Lambda ^{\prime }\overline{V}%
_{\Lambda ^{\prime }}\circ G_{\Lambda ^{\prime }}\circ S_{\Lambda ^{\prime
}}\circ \overline{V}_{\Lambda ^{\prime }}  \notag \\
&&\left. +S_{\Lambda }\circ \int\limits_{\Lambda }^{\Lambda _{0}}d\Lambda
^{\prime }T_{\Lambda ^{\prime }}\circ G_{\Lambda ^{\prime }}\circ \overline{V%
}_{\Lambda ^{\prime }}\circ G_{\Lambda ^{\prime }}\circ \overline{V}%
_{\Lambda ^{\prime }}\circ S_{\Lambda ^{\prime }}\right\}
\end{eqnarray}%
with the initial condition $\chi _{\Lambda _{0}}=0$ and $T_{\Lambda _{0}}$
is determined by the symmetry of the order parameter, e.g. $T_{\Lambda
_{0}}=1$ for the antiferromagnetic and singlet superconducting
susceptibility, $T_{\Lambda _{0}}(k)=\cos k_{y}-\cos k_{x}$ for the d-wave
superconducting susceptibility etc.

Let consider the local in $\Lambda $ version of the equations (\ref{TwoLoopA}%
), which is obtained by the replacement $\overline{V}_{\Lambda ^{\prime
}}\rightarrow \overline{V}_{\Lambda },$ $\Sigma _{\Lambda ^{\prime
}}\rightarrow \Sigma _{\Lambda }$. This replacement introduces corrections
of the same order, which are already neglected in the two-loop
approximation, and therefore, can be considered on the same level of an
approximation. In this way we obtain 
\end{subequations}
\begin{subequations}
\label{TwoLoopLoc}
\begin{eqnarray}
\frac{d\Sigma _{\Lambda }}{d\Lambda } &=&\overline{V}_{\Lambda }\circ
S_{\Lambda }+S_{\Lambda }\circ \overline{V}_{\Lambda }\circ \left[ (1-%
\widehat{\mathcal{P}})(G_{\Lambda }\circ G_{\Lambda })\right] \circ 
\overline{V}_{\Lambda }  \label{TwoLoopLoc1} \\
\frac{d\overline{V}_{\Lambda }}{d\Lambda } &=&\widehat{\mathcal{P}}\left\{ 
\overline{V}_{\Lambda }\circ \frac{d}{d\Lambda }(G_{\Lambda }\circ
G_{\Lambda })\circ \overline{V}_{\Lambda }\right.  \notag \\
&&+\overline{V}_{\Lambda }\circ (G_{\Lambda }\circ S_{\Lambda })\circ 
\overline{V}_{\Lambda }\circ \left[ (1-\widehat{\mathcal{P}})(G_{\Lambda
}\circ G_{\Lambda })\right] \circ \overline{V}_{\Lambda }  \notag \\
&&\left. +S_{\Lambda }\circ \overline{V}_{\Lambda }\circ G_{\Lambda }\circ 
\overline{V}_{\Lambda }\circ G_{\Lambda }\circ \overline{V}_{\Lambda }\circ
G_{\Lambda }\right\}
\end{eqnarray}%
The local equations (\ref{TwoLoopLoc}) have a similar form as the two-loop
equations in the field-theoretical approaches, e.g. for the 1D fermionic
systems \cite{Solyom}, and, therefore, can be used to make connection with
these approaches. The terms with the projection operator $\widehat{\mathcal{P%
}}$ coming from Eq. (\ref{Sub}) substract the one-loop ($\ln ^{2}$ in 1D
case) contributions from the third-order diagrams for the vertex. The
coresponding contribution to the self-energy (last term in the first
equation) is a frequency-independent constant, which can be omitted.
Contrary to field-theoretical approaches, the Eqs. (\ref{TwoLoopLoc})
account for both, regular and singular terms in the perturbation expansion
and are written for the coupling constants themselves, not for their
invariant combinations with the self-energy.

\bigskip To solve numerically Eqs. (\ref{TwoLoopA}), (\ref{TwoLoopHi}), or (%
\ref{TwoLoopLoc}), we use the discretization of momentum space in $2$
patches (L and R) in one dimension and $N_{p}=24$ patches in two dimensions
with the same patching scheme as proposed in Ref. \cite{SalmHon1}. For the
frequency dependence of the self-energy we use $N_{\omega }=100$ frequencies
i$\nu _{i}$ suitably chosen on the imaginary axis (these frequencies do not
have to coincide with the Matsubara frequencies since for a
frequency-independent $\overline{V}$ the self-energy is defined on the
entire imaginary axis, cf. Ref. \cite{KK}). We account for the self-energy
effects by expanding the self-energy $\Sigma _{\Lambda }(i\nu )$ around $\nu
=0$ and introducing $Z$-factors 
\end{subequations}
\begin{equation}
Z_{\mathbf{k}_{F}}^{\Lambda }=[1-\partial \mathrm{Im}\Sigma _{\Lambda }(%
\mathbf{k}_{F},i\nu )/\partial \nu ]_{\nu =0}^{-1},  \label{Z}
\end{equation}%
the feedback of the imaginary part of the self-energy to the flow of
vertices is neglected. The Green functions in this approximation take the
form \cite{SalmHonSE}%
\begin{eqnarray}
G_{\Lambda }(\mathbf{k},i\nu _{n}) &=&\frac{Z_{\mathbf{k}_{F}}^{\Lambda
}\theta (|\varepsilon _{\mathbf{k}}|-\Lambda )}{i\nu _{n}-\varepsilon _{%
\mathbf{k}}}  \label{GZ} \\
S_{\Lambda }(\mathbf{k},i\nu _{n}) &=&-\frac{Z_{\mathbf{k}_{F}}^{\Lambda
}\delta (|\varepsilon _{\mathbf{k}}|-\Lambda )}{i\nu _{n}-\varepsilon _{%
\mathbf{k}}}  \notag
\end{eqnarray}%
where $\mathbf{k}_{F}$ corresponds to the projection of the vector $\mathbf{k%
}$ to the Fermi surface.

The derivative of the self-energy which enters the Eq. (\ref{Z}) is
determined numerically from the values of the self-energy at the first two
frequencies i$\nu _{1,2}.$ The approximation (\ref{GZ}) can be applied only
in the paramagnetic state without strong exchange and/or umklapp scattering
(i.e. away from half filling in one dimension and at not too low
temperatures and not too close to the van Hove band fillings in two
dimensions). Above the antiferromagnetically ordered ground state the
divergence of the corresponding vertices leads to a pseudogap structure of
the self-energy and spectral functions \cite{KK,Metz}. This structure can be
correctly described only with the frequency-dependent self energy and is not
considered here. We also neglect the first and third terms in the flow
equations for the self-energy (\ref{TwoLoopA}) and (\ref{TwoLoopLoc}) as
responsible purely for the deformation of the Fermi surface by the
interaction. This deformation was found numerically to be small in two
dimensions at small next-nearest hopping $t^{\prime }$ \cite{SalmHon} and
can be treated accurately by introducing corresponding counterterms\cite%
{SalmBook,Metz1,SalmhofSE}.

\section{Results}

\subsection{1D case}

First we consider the results for the 1D electronic dispersion. 
\begin{equation}
\varepsilon _{k}=-2t\cos k-\mu  \label{ek1D}
\end{equation}%
In this case we have only 2 patches (L and R) at $k_{F}=\pm \arccos (\mu /2)$%
. After the projection to the Fermi points, only 4 independent vertices
remain: $V_{1}=V(L,R;R,L),$ $V_{2}=V(L,R;L,R),V_{3}=V(L,L;R,R),$ and $%
V_{4}=V(L,L;L,L)=V(R,R;R,R).$ With the linearization of the dispersion (\ref%
{ek1D}) near the Fermi points, the flow of these vertices in the two-loop
approximation is well-studied in the field-theoretical approach\cite{Solyom}
and it is described by the equations:%
\begin{eqnarray}
dg_{1}/dl &=&\frac{1}{\pi v_{F}}g_{1}^{2}+\frac{1}{2\pi ^{2}v_{F}^{2}}%
g_{1}^{2}(g_{1}+g_{4})  \label{SolyomEq} \\
dg_{2}^{\prime }/dl &=&\frac{1}{\pi v_{F}}g_{3}^{2}+\frac{1}{2\pi
^{2}v_{F}^{2}}g_{3}^{2}(g_{1}-2g_{2}-g_{4})  \notag \\
dg_{3}/dl &=&\frac{1}{\pi v_{F}}g_{2}^{\prime }g_{3}+\frac{1}{4\pi
^{2}v_{F}^{2}}g_{3}[(g_{2}^{\prime })^{2}+g_{3}^{2}-2g_{2}^{\prime }g_{4}] 
\notag \\
dg_{4}/dl &=&\frac{3}{4\pi ^{2}v_{F}^{2}}(g_{2}^{\prime }g_{3}^{2}-g_{1}^{3})
\notag \\
d\ln Z/dl &=&\frac{1}{4\pi ^{2}v_{F}^{2}}%
(g_{1}^{2}-g_{1}g_{2}+g_{2}^{2}+g_{3}^{2})
\end{eqnarray}%
where $g_{i}=Z^{2}V_{i}$ are the invariant coupling constants, $%
g_{2}^{\prime }=g_{1}-2g_{2}$, $v_{F}=2t\sin k_{F}$, $l=\ln \Lambda $. We
emphasize once more, that the difference of the equations (\ref{TwoLoopLoc})
and (\ref{SolyomEq}) is that the latter account for the universal
contributions to the flow of the coupling constants only, while the former
treat also the non-universal contributions, e.g. connected with the
nonlinearity of the dispersion.

The result of the solution of Eqs. (\ref{TwoLoopLoc}) for $g_{1}=t,$ $%
g_{2}=2t,$ $g_{3}=g_{4}=\mu =0,$ and $T=10^{-4}t$ is presented in Fig. 4.
Chosing $g_{3}=0$ allows us to avoid the effect of umklapp scattering. To
verify that the result of the solution of the Eqs. (\ref{SolyomEq}) is
indeed reproduced at $\Lambda \ll 1$, we use the result of integration of
Eqs. (\ref{TwoLoopLoc}) at some cutoff parameter $\Lambda _{c}=e^{-4}t\ll 1$
as a starting condition for the Eqs. (\ref{SolyomEq}) and compare the result
of the solution of the Eqs. (\ref{TwoLoopLoc}) and (\ref{SolyomEq}) for $%
\Lambda <\Lambda _{c}$. One can see that the results of the local two-loop
fRG approach agree with the corresponding results of the field-theoretical
approach, Eqs. (\ref{SolyomEq}). At the same time, the results of the
solution of local and nonlocal fRG equations are different due to the
nonuniversal initial part of the flow. We have verified that this is mainly
connected with the momentum dependence of the Fermi velocity, the difference
almost disappears for the linearized version of the dispersion (\ref{ek1D}).
The nonlocal equations (\ref{TwoLoopA1}) are expected to treat better the
effect of the nonlinearity of the dispersion, therefore we consider only
their solution in the 2D case below.

\subsection{2D case}

For the discussion of the results of fRG approach in 2D case we also
consider the solution of the Eqs. (\ref{TwoLoopA}) without the two-loop
corrections (i.e. without the last term in the second equation) to
investigate how the one-loop flow changes due to correction of the errors of
vertex projections by the second and third term in the Eq. (\ref{TwoLoopA2}%
). The difference of the latter results from the one-loop results shows the
effect of the vertex projection on the renormalization group flows, while
their difference to the two-loop results shows the effect of the two-loop
corrections.

First we consider the results of the solution of Eqs. (\ref{TwoLoopA}) for
the dispersion 
\begin{equation}
\varepsilon _{\mathbf{k}}=-t(\cos p_{x}+\cos p_{y})+t|\cos p_{x}-\cos
p_{y}|-\mu  \label{ekf}
\end{equation}%
The corresponding Fermi surface has flat parts along the directions $%
|p_{x}|,|p_{y}|=\arccos (-\mu /2).$ The field-theoretical approach for a
flat Fermi surface was applied earliar in Ref. \cite{Freire,Freire1}. The
results of the numerical solution of Eqs. (\ref{TwoLoopA}) for $\mu =t,$ $%
U=7.81t$ are shown in Fig. 5. We choose this relatively large value of the
interaction $U$ since it corresponds to the value of the dimensionless
coupling constant $U\Delta /(\pi v_{F})=3$ used in Refs. \cite%
{Freire,Freire1} where $v_{F}$ is the Fermi velocity, $\Delta $ is the
length of the Fermi surface flat parts ($v_{F}=\sqrt{4t^{2}-\mu ^{2}},$ $%
\Delta =2\arccos (-\mu /(2t))$ for the dispersion (\ref{ekf})); a larger
value $U\Delta /(\pi v_{F})=10$ was considered in Ref. \cite{Freire}. One
can see that the vertices without the two-loop corrections diverge at much
larger energy scales compared to the two-loop results in agreeement with
Refs. \cite{Freire,Freire1}. The scale of the vertex divergence in the
one-loop calculation with partly corrected projection errors (Eqs. (\ref%
{TwoLoopA}) withot the last term in the second equation) agrees with the
result of the one-loop field-theoretical approach (not having any projection
errors), but is larger than the scale of the vertex divergence in one-loop
approach with vertex projection. The flow of the vertices in the two-loop
approach agrees also with the results of Refs. \cite{Freire,Freire1}.

The functional renormalization group approach can be further applied to the $%
t$-$t^{\prime }$ Hubbard model with the dispersion%
\begin{equation}
\varepsilon _{\mathbf{k}}=-2t(\cos p_{x}+\cos p_{y})+4t^{\prime }(\cos
p_{x}\cos p_{y}+1)-\mu
\end{equation}%
where the conventional field-theoretical approach is not applicable due to
the presence of squared logarithmic singularities in the perturbation series
near van Hove band filling ($\mu =0$). The results for the flow of vertices
and susceptibilities for $t^{\prime }=0.1t,$ $U=2t$ and the fillings close
to vH band filling ($\mu =\pm 0.1t$) are presented in Fig.6. At $\mu =0.1t$
(above vH filling) the largest susceptibility is observed with respect to
the antiferromagnetic instability in both, the one- and the two-loop
approaches. For this value of $\mu $ and chosen temperature $T=0.1t$ the
results for the susceptibilities in one- and two-loop approach substantially
differ. The antiferromagnetic susceptibility in the end of the flow in the
one-loop approach with partly corrected errors of vertex projection (Eqs. (%
\ref{TwoLoopA}) with omitted last term in the second equation) is larger
than the results of this approach with vertex projection and deviates more
from the two-loop results. Therefore, the results of one-loop approach with
vertex projection agree better with the two-loop results, which is possibly
due to account of only universal terms of the flow in these approaches. For $%
\mu =-0.1t$ (below vH filling) we observe the same qualitative tendencies
with smaller difference of the results of one- and two-loop approaches. With
decreasing temperature, the superconducting instability becomes dominating
in this case (see below).

The results for the flow of the vertices and susceptibilities at the filling
further from vH one ($\mu =-0.5t$) are shown in Fig. 7. At this filling and
not too low temperatures the antiferromagnetic susceptibility is the largest
one (not shown), but with decreasing temperature the d-wave superconducting
instability becomes the leading instability. Susceptibilities in the
one-loop approach with partial correction of vertex projection errors are
larger than the one- and the two-loop approaches. The susceptibilities in
the end of the flow of one- and two-loop approaches are, however, closer to
each other than in the above considered case $\mu =-0.1t$.

The calculated temperature dependences of the susceptibilities for
antiferromagnetic and superconducting instabilities, as well, as Z-factors
for $\mu =\pm 0.1t$ are shown in Fig. 8. At $\mu =0.1t$ we observe a maximum
of the antiferromagnetic susceptibility in the two-loop approach, while the
corresponding susceptibility in the one-loop approach diverges with
decreasing temperature (Fig. 8a). More generally, we find that the
divergence of the vertices (and susceptibilities) is strongly suppressed in
the two-loop approach. This divergence is not, however, fully removed, since
for smaller $\mu <0.08$ we find again the possibility of the
antiferromagnetic ground state in the two-loop approach. At $\mu =-0.1t$ the
antiferromagnetic susceptibility also has a maximum at some temperature and
then decreases with decreasing $T$ (Fig. 8b), while the superconducting
susceptibility increases, showing the possibility of the superconducting
ground state in both, one- and two-loop approaches. The increase of $\chi _{%
\text{dSC}}$ in the one-loop approach is again more pronounced than in the
two-loop approach, so that the temperature where the susceptibility diverges
in the two-loop approach is expected to be much smaller than in the one-loop
approach. The $Z$-factors decrease almost linearly with $\ln (t/T)$ at
intermediate temperatures, but below the temperature, where the maximum of
the susceptibility is reached, their temperature dependence becomes linear
in $T$ and therefore $Z_{\mathbf{k}_{F}}$ are not expected therefore to
vanish at lower temperatures. We have also verified during the calculations
that the imaginary part of the self-energy remains linear function of $\nu $
at small imaginary frequencies.

In Fig. 9 we summarize the results for the qp damping $\Gamma _{\mathbf{k}%
_{F}}=-\mathrm{Im}\Sigma (\mathbf{k}_{F},0)$ and the FS shift $\mathrm{Re}%
\Sigma (\mathbf{k}_{F},0)$ estimated at different FS points in the end of
the two-loop fRG flow. The qp damping depends almost linearly on temperature
at not too low temperatures; this dependence becomes quadratic at low $T$.
The observed linear dependence of the scattering rates at not too low
temperatures may be due to closeness to the antiferromagnetic quantum
critical point; more detail investigations of this dependence are, however,
required. At the same time, the quadratic temperature dependence at low
temperatures supports the Fermi-liquid picture in this temperature range
above the paramagnetic ground state. The Fermi surface shift contributions
are small and negative for $\mu <0$. For $\mu >0$ the Fermi surface shifts
have opposite signs at the point closest to the ($\pi ,0$) point and to the
diagonal, leading therefore to small deformation of the Fermi surface, which
makes it flatter.

\section{Discussion and conclusions}

We have considered the effect of the two-loop corrections on the fRG flow.
In 1D case we find that the nonuniversal corrections contribute to the flow
at large momenta scales, while at small momentum scales we have recovered
the results obtained from the field-theoretical RG approach. For the 2D case
with flat Fermi surface we also find good agreement with the previous
results of the field-theoretical RG. The fRG approach was applied further to
the case of curved Fermi surface without nesting, where we obtained the flow
of the vertices and susceptibilities at the two-loop level.

In two-dimensions the two-loop corrections do not change the leading
instability, but may lead to slight shift of the phase boundaries in
comparison with the previous one-loop analysis. The difference of the
two-loop results and the one-loop results with projected vertices in two
dimensions is smaller than to the one-loop results with partly corrected
projection errors and decreases going away from the van Hove band filling.
Therefore, the commonly used one-loop fRG approach with projected vertices
serves as a good starting point for calculating higher-loop corrections.

We have also considered the flow of the qp spectral weight, qp damping and
the Fermi surface shift. In agreement with earliar studies, in the
considered cases of not too low temperatures in two dimensions we obtain the
qp weight $Z\simeq 0.9$, so that the quasiparticles remain well defined
during the fRG flow. The qp damping and estimated Fermi surface shifts are
also numerically small.

Possible future applications of the method would be its implementation
within the temperature-cutoff fRG scheme\cite{SalmHon1}, where the two-loop
corrections are expected to be smaller than for momentum cutoff due to
better treatment of degrees of freedoom with different excitation energy.
The calculation of the two-loop corrections for the temperature cutoff is,
however, more difficult task since it requries more intense numerical
calculations caused by the smoothness of the cutoff. Another possible
extension of the method would include consideration of the frequency
dependence of the self-energy and/or vertices, which also has to be
performed.

\section{Acknowledgements}

We are grateful to W. Metzner for stimulating discussions and a careful
reading of the manuscript. This work is supported in part by Grant No.
4640.2006.2 (Support of Scientific Schools) and 07-02-01264a from the
Russian Basic Research Foundation.

\section{Appendix. Derivation of the two-loop fRG equations}

In this Appendix we consider the derivation of the two-loop RG equations. We
use the notations of Refs. \cite{SalmHon,SalmHonR}, which considered the 1PI
RG equations for the terms of the expansion of the 1PI generating functional
in fermionic field,%
\begin{equation*}
\Gamma _{\Lambda }(\phi )=\sum_{m\geq 0}\gamma _{\Lambda }^{(m)}(\phi )
\end{equation*}%
where $\phi (X)$ are fermionic fields, 
\begin{equation}
X=(x,\tau ,\sigma ,\pm )
\end{equation}%
is the short notation for the space, time, spin, and charge variables (the $%
\pm $ sign corresponds to the incoming and the outgoing particles,
respectively). At the two-loop order the hierarchy of RG equations for the
1PI functions is truncated at the three-particle vertex and has the form 
\cite{SalmHon,SalmHonR}%
\begin{eqnarray}
\overset{.}{\gamma }_{\Lambda }^{(2)}(\phi ) &=&(\phi ,Q\phi )+\frac{1}{2}%
\text{Tr}[S_{\Lambda }\widetilde{\gamma }_{\Lambda }^{(2)}]  \notag \\
\overset{.}{\gamma }_{\Lambda }^{(4)}(\phi ) &=&\frac{1}{2}\text{Tr}%
[S_{\Lambda }\widetilde{\gamma }_{\Lambda }^{(4)}]-\frac{1}{2}\text{Tr}%
[S_{\Lambda }\widetilde{\gamma }_{\Lambda }^{(2)}G_{\Lambda }\widetilde{%
\gamma }_{\Lambda }^{(2)}]  \notag \\
\overset{.}{\gamma }_{\Lambda }^{(6)}(\phi ) &=&-\frac{1}{2}\text{Tr}%
[S_{\Lambda }\widetilde{\gamma }_{\Lambda }^{(4)}G_{\Lambda }\widetilde{%
\gamma }_{\Lambda }^{(2)}+S_{\Lambda }\widetilde{\gamma }_{\Lambda
}^{(2)}G_{\Lambda }\widetilde{\gamma }_{\Lambda }^{(4)}]+\frac{1}{2}\text{Tr}%
[S_{\Lambda }\widetilde{\gamma }_{\Lambda }^{(2)}G_{\Lambda }\widetilde{%
\gamma }_{\Lambda }^{(2)}G_{\Lambda }\widetilde{\gamma }_{\Lambda
}^{(2)}]+O(\gamma _{\Lambda }^{(8)})  \label{RG}
\end{eqnarray}%
where 
\begin{equation}
\widetilde{\gamma }_{\Lambda }^{(m)}(X,Y,\phi )=\frac{\delta }{\delta \phi
(X)}\frac{\delta }{\delta \phi (Y)}\gamma _{\Lambda }^{(m+2)}(\phi )
\end{equation}%
the trace is taken with respect to $X$-variables and dots denote derivatives
with respect to $\Lambda $.

For practical calculations the vertices $\gamma _{\Lambda }^{(m)}(\phi )$
are expressed explicitly through the $\phi $-fields as 
\begin{equation}
\gamma _{\Lambda }^{(m)}(\phi )=\frac{1}{m!}\sum_{m}\int dX^{m}\gamma
_{m}(\Lambda |X)\phi (X_{1})...\phi (X_{m})
\end{equation}%
and 
\begin{equation}
\widetilde{\gamma }_{\Lambda }^{(m)}(X,Y,\phi )=\frac{1}{m!}\int
d^{m}X^{\prime }\gamma _{m+2}(\Lambda |X,Y,X^{\prime })\phi (X_{1}^{\prime
})...\phi (X_{m}^{\prime })
\end{equation}%
In these notations the RG equations (\ref{RG})\ read 
\begin{subequations}
\label{TwoLoop}
\begin{eqnarray}
\overset{.}{\gamma }_{2}(\Lambda |X) &=&\frac{1}{2}\int d^{2}Y\gamma
_{4}(\Lambda |X,Y)S_{\Lambda }(Y)  \label{TL1} \\
\overset{.}{\gamma }_{4}(\Lambda |X) &=&\frac{1}{2}\int d^{2}Y\gamma
_{6}(\Lambda |X,Y)S_{\Lambda }(Y)-\frac{1}{2}\int d^{4}YB_{\Lambda
}(X,Y)L_{\Lambda }(Y)  \label{TL2} \\
\overset{.}{\gamma }_{6}(\Lambda |X) &=&\frac{1}{2}\int d^{6}YD_{\Lambda
}(X,Y)M_{\Lambda }(Y)-\frac{1}{2}\int d^{4}YE_{\Lambda }(X,Y)L_{\Lambda }(Y)
\label{TL3}
\end{eqnarray}%
where 
\end{subequations}
\begin{eqnarray}
L_{\Lambda }(Y) &=&S_{\Lambda }(Y_{1},Y_{2})G_{\Lambda
}(Y_{3},Y_{4})+G_{\Lambda }(Y_{1},Y_{2})S_{\Lambda }(Y_{3},Y_{4}) \\
B_{\Lambda }(X,Y) &=&\gamma _{4}(\Lambda |X_{1}X_{2};Y_{2},Y_{3})\gamma
_{4}(\Lambda |X_{3}X_{4};Y_{4},Y_{1})  \notag \\
&&-\gamma _{4}(\Lambda |X_{1}X_{3};Y_{2},Y_{3})\gamma _{4}(\Lambda
|X_{2}X_{4};Y_{4},Y_{1})  \notag \\
&&+\gamma _{4}(\Lambda |X_{1}X_{4};Y_{2},Y_{3})\gamma _{4}(\Lambda
|X_{2}X_{3};Y_{4},Y_{1})
\end{eqnarray}%
and 
\begin{eqnarray}
M_{\Lambda }(Y) &=&S_{\Lambda }(Y_{1},Y_{2})G_{\Lambda
}(Y_{3},Y_{4})G_{\Lambda }(Y_{5},Y_{6})  \notag \\
&&+G_{\Lambda }(Y_{1},Y_{2})S_{\Lambda }(Y_{3},Y_{4})G_{\Lambda
}(Y_{5},Y_{6})  \notag \\
&&+G_{\Lambda }(Y_{1},Y_{2})G_{\Lambda }(Y_{3},Y_{4})S_{\Lambda
}(Y_{5},Y_{6}) \\
D_{\Lambda }(X,Y) &=&\gamma _{4}(\Lambda |X_{1},X_{2};Y_{2}Y_{3})\gamma
_{4}(\Lambda |X_{3}X_{4};Y_{4}Y_{5})\gamma _{4}(\Lambda
|X_{5}X_{6};Y_{6}Y_{!})  \notag \\
&&+14\text{ permutations} \\
E_{\Lambda }(X,Y) &=&\gamma _{4}(\Lambda |X_{1}X_{2};Y_{2}Y_{3})\gamma
_{6}(\Lambda |X_{3}X_{4}X_{5}X_{6};Y_{4}Y_{1})  \notag \\
&&+14\text{ permutations}
\end{eqnarray}

As discussed in main text, in the present paper we consider the expansion in
the number of loops rather than in effective vertices. Iterating the Eq. (%
\ref{TL3}) for $\gamma ^{(6)},$ one can easily see that the first term in
this equation corresponds to the three-loop contribution to $\gamma ^{(4)}.$
Therefore, in the following we neglect this term. Due to this neglection,
the Eq. (\ref{TL3})\ can be integrated analytically. Substituting the result
into Eq. (\ref{TL2}) we obtain 
\begin{eqnarray}
\overset{.}{\gamma }_{2}(\Lambda |X) &=&\frac{1}{2}\int d^{2}Y\gamma
_{4}(\Lambda |X,Y)S_{\Lambda }(Y)  \notag \\
\overset{.}{\gamma }_{4}(\Lambda |X) &=&\frac{1}{4}\int\limits_{\Lambda
}^{\Lambda _{0}}d\Lambda ^{\prime }\int d^{2}Y\int d^{6}Y^{\prime
}S_{\Lambda }(Y)M_{\Lambda }(Y^{\prime })D_{\Lambda }((X,Y),Y^{\prime }) 
\notag \\
&&-\frac{1}{2}\int d^{4}YL_{\Lambda }(Y)B_{\Lambda }(X,Y)
\end{eqnarray}%
This substitution considerebly simplifies the numerical solution of the
equations since one has to consider the vertices $\gamma _{2}$ and $\gamma
_{4}$ only.

Introducing the self-energy and $2$-particle irreducible vertex by 
\begin{eqnarray}
\Sigma (\xi _{1},\xi _{2}) &=&\gamma _{2}((\xi _{1},\uparrow ,+),(\xi
_{2},\uparrow ,-))  \notag \\
V(\xi _{1}\xi _{2};\xi _{3}\xi _{4}) &=&\gamma _{4}((\xi _{1},\uparrow
,+),(\xi _{2},\downarrow ,+),(\xi _{3},\uparrow ,-),(\xi _{4},\downarrow ,-))
\end{eqnarray}%
where $\xi _{i}=(\mathbf{x}_{i},\tau _{i})$ and the Fourier transformed
quantities 
\begin{eqnarray}
\Sigma (k) &=&\int d^{2}\xi \Sigma (\xi _{1},\xi _{2})e^{i(\xi _{1}-\xi
_{2})k}  \notag \\
V(k_{1}k_{2};k_{3}k_{4}) &=&\int d^{4}\xi V(\xi _{1}\xi _{2};\xi _{3}\xi
_{4})e^{i\xi _{1}k_{1}+i\xi _{2}k_{2}-i\xi _{3}k_{3}-i\xi _{4}k_{4}}
\end{eqnarray}%
where $k_{i}=(\mathbf{k}_{i},i\omega _{n}^{(i)})$ and exploiting charge-,
spin- and translational invariance in the same way as in Refs. \cite%
{SalmHon,SalmHonR} we obtain the equations (\ref{TwoLoop})\ of the paper.

Similar derivation can be performed for susceptibilities. Following Ref. 
\cite{SalmHonR} we introduce vertices $\gamma _{\Lambda }^{(m,n)}$ where $m$
refers to the number of boson lines and $n$ to the number of fermion lines,
which enter or go out of the vertex. The equations for the vertices $\gamma
_{\Lambda }^{(m,n)}$ read%
\begin{eqnarray}
\overset{.}{\gamma }_{\Lambda }^{(2,0)}(\phi ) &=&\frac{1}{2}\text{Tr}%
[S_{\Lambda }\widetilde{\gamma }_{\Lambda }^{(1,0)}G_{\Lambda }\widetilde{%
\gamma }_{\Lambda }^{(1,0)}G_{\Lambda }] \\
\overset{.}{\gamma }_{\Lambda }^{(1,2)}(\phi ) &=&-\frac{1}{2}\text{Tr}%
[S_{\Lambda }\widetilde{\gamma }_{\Lambda }^{(1,2)}]+\frac{1}{2}\text{Tr}%
[S_{\Lambda }\widetilde{\gamma }_{\Lambda }^{(1,0)}G_{\Lambda }\widetilde{%
\gamma }_{\Lambda }^{(0,2)}] \\
\overset{.}{\gamma }_{\Lambda }^{(1,4)}(\phi ) &=&-\frac{1}{2}\text{Tr}%
[S_{\Lambda }\widetilde{\gamma }_{\Lambda }^{(1,4)}]+\frac{1}{2}\text{Tr}%
[S_{\Lambda }\widetilde{\gamma }_{\Lambda }^{(1,0)}G_{\Lambda }\widetilde{%
\gamma }_{\Lambda }^{(0,4)}]+\frac{1}{2}\text{Tr}[S_{\Lambda }\widetilde{%
\gamma }_{\Lambda }^{(1,2)}G_{\Lambda }\widetilde{\gamma }_{\Lambda
}^{(0,2)}]  \notag \\
&&+\frac{1}{2}\text{Tr}[S_{\Lambda }\widetilde{\gamma }_{\Lambda
}^{(1,0)}G_{\Lambda }\widetilde{\gamma }_{\Lambda }^{(0,2)}G_{\Lambda }%
\widetilde{\gamma }_{\Lambda }^{(0,2)}]
\end{eqnarray}%
Expanding again $\gamma ^{(m,n)}(\phi )$ in $\phi $, neglecting first three
terms in the equation for $\overset{.}{\gamma }_{\Lambda }^{(1,4)}(\phi )$
as corresponding to the higher-loop order, substituting the result for $%
\overset{.}{\gamma }_{\Lambda }^{(1,4)}(\phi )$ in the equation for $\overset%
{.}{\gamma }_{\Lambda }^{(1,2)}(\phi ),$ we obtain%
\begin{eqnarray}
\overset{.}{\gamma }^{(1,0)}(\Lambda |X,X^{\prime }) &=&\int
d^{4}YL_{\Lambda }(Y)\gamma ^{(1,2)}(\Lambda |X,Y_{2},Y_{3})\gamma
^{(1,2)}(\Lambda |X^{\prime },Y_{4},Y_{1}) \\
\overset{.}{\gamma }^{(1,2)}(\Lambda |X,X^{\prime }) &=&\frac{1}{4}%
\int\limits_{\Lambda }^{\Lambda _{0}}d\Lambda ^{\prime }\int d^{2}Y\int
d^{6}Y^{\prime }S_{\Lambda }(Y)M_{\Lambda }(Y^{\prime })\widetilde{D}%
_{\Lambda }(X,(X^{\prime },Y),Y^{\prime })  \notag \\
&&-\frac{1}{2}\int d^{4}YL_{\Lambda }(Y)\widetilde{B}_{\Lambda }(X,X^{\prime
},Y)
\end{eqnarray}%
where 
\begin{eqnarray}
\widetilde{B}_{\Lambda }(X,X^{\prime },Y) &=&\gamma _{(1,2)}(\Lambda
|X;Y_{2},Y_{3})\gamma _{4}(\Lambda |X_{1}^{\prime }X_{2}^{\prime
};Y_{4},Y_{1}) \\
\widetilde{D}_{\Lambda }(X,X^{\prime },Y) &=&\gamma _{(1,2)}(\Lambda
|X;Y_{2}Y_{3})\gamma _{4}(\Lambda |X_{1}^{\prime }X_{2}^{\prime
};Y_{4}Y_{5})\gamma _{4}(\Lambda |X_{3}^{\prime }X_{4}^{\prime };Y_{6}Y_{!})
\notag \\
&&+\text{permutations (}X^{\prime }\text{)}
\end{eqnarray}

\begin{figure}[tbp]
\includegraphics[width=14cm]{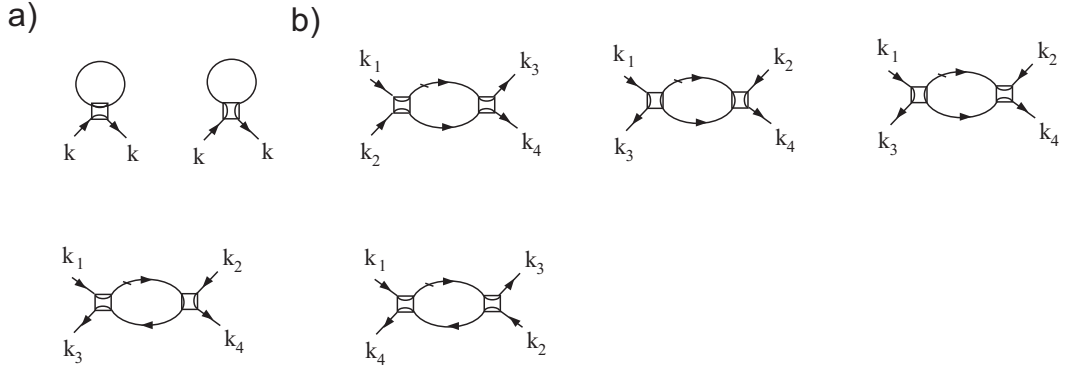} 
\caption{ The diagrams for the self-energy (a) and vertex (b) flow at the
one-loop order. The solid lines correspond to the cut propagator $G_{\Lambda
}$, the lines with dash - to the single-scale propagator $S_{\Lambda },$
boxes - to the vertices $V_{\Lambda }.$ Lines inside the box show the
direction of spin conservation. }
\end{figure}

\begin{figure}[tbp]
\includegraphics[width=8cm]{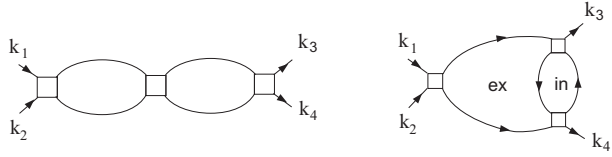} 
\caption{ Ladder-type (left) and non-ladder (right) diagrams in the third
order perturbation theory. Solid lines correspond to the bare electronic
prpagator $G_{0},$ boxes - to the bare interaction $U,$ "in" \ and "ex"
denote internal and external bubble in the non-ladder diagrams.}
\end{figure}

\begin{figure}[tbp]
\includegraphics[width=12cm]{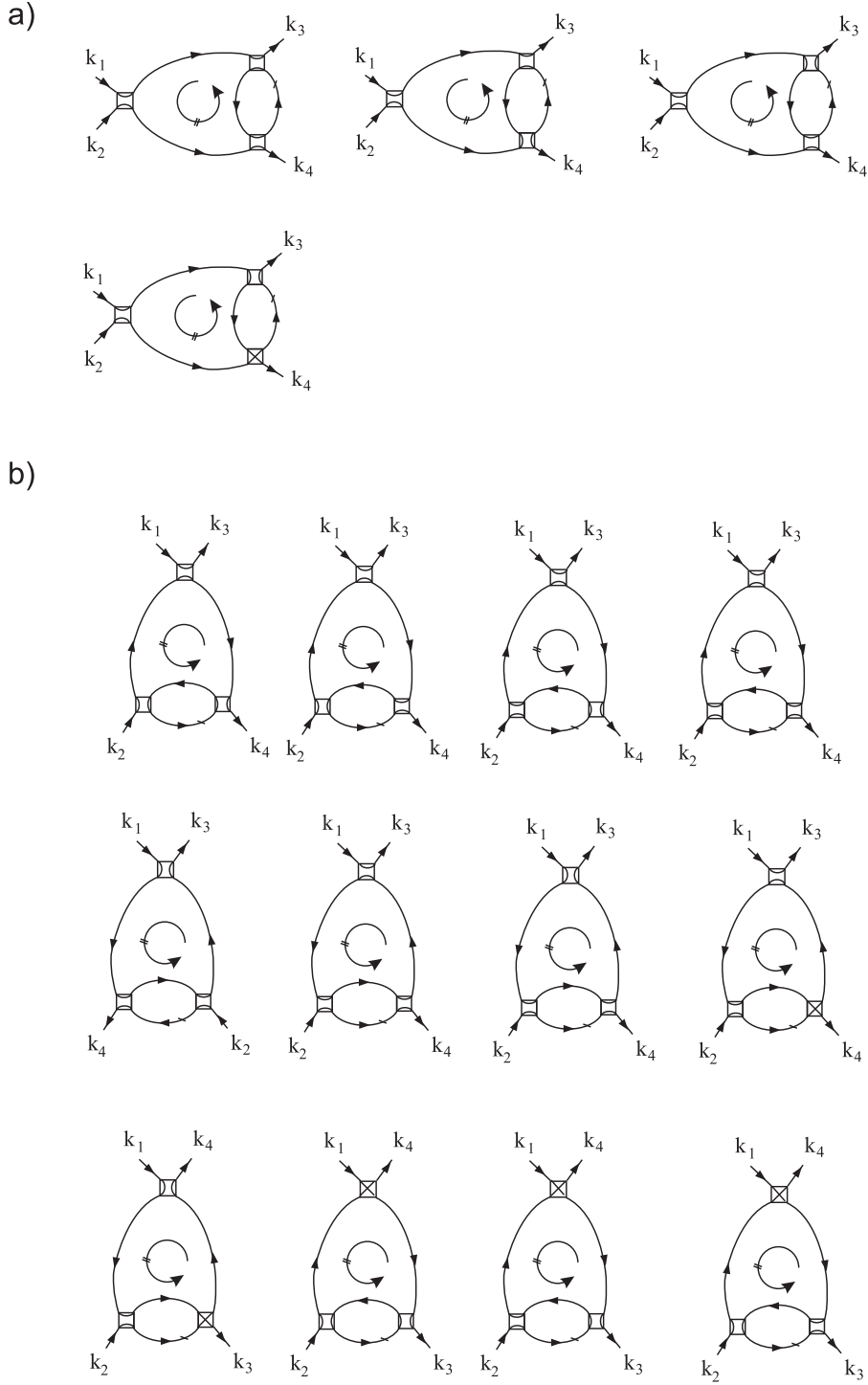} 
\caption{ The diagrams for the contributions to the flow of the vertex at
the two-loop order in the particle-particle (a) and the particle-hole (b)
channels. The two of the three lines without dash correspond to the $%
G_{\Lambda ^{\prime }}$ propagator and one to the $S_{\Lambda ^{\prime }}$
propagator (the circle arrow with double dash denotes their permutations),
the line with dash corresponds to the single-scale propagator $S_{\Lambda }.$
The other notations are the same as in Fig. 1.}
\end{figure}

\begin{figure}[tbp]
\includegraphics[width=8cm]{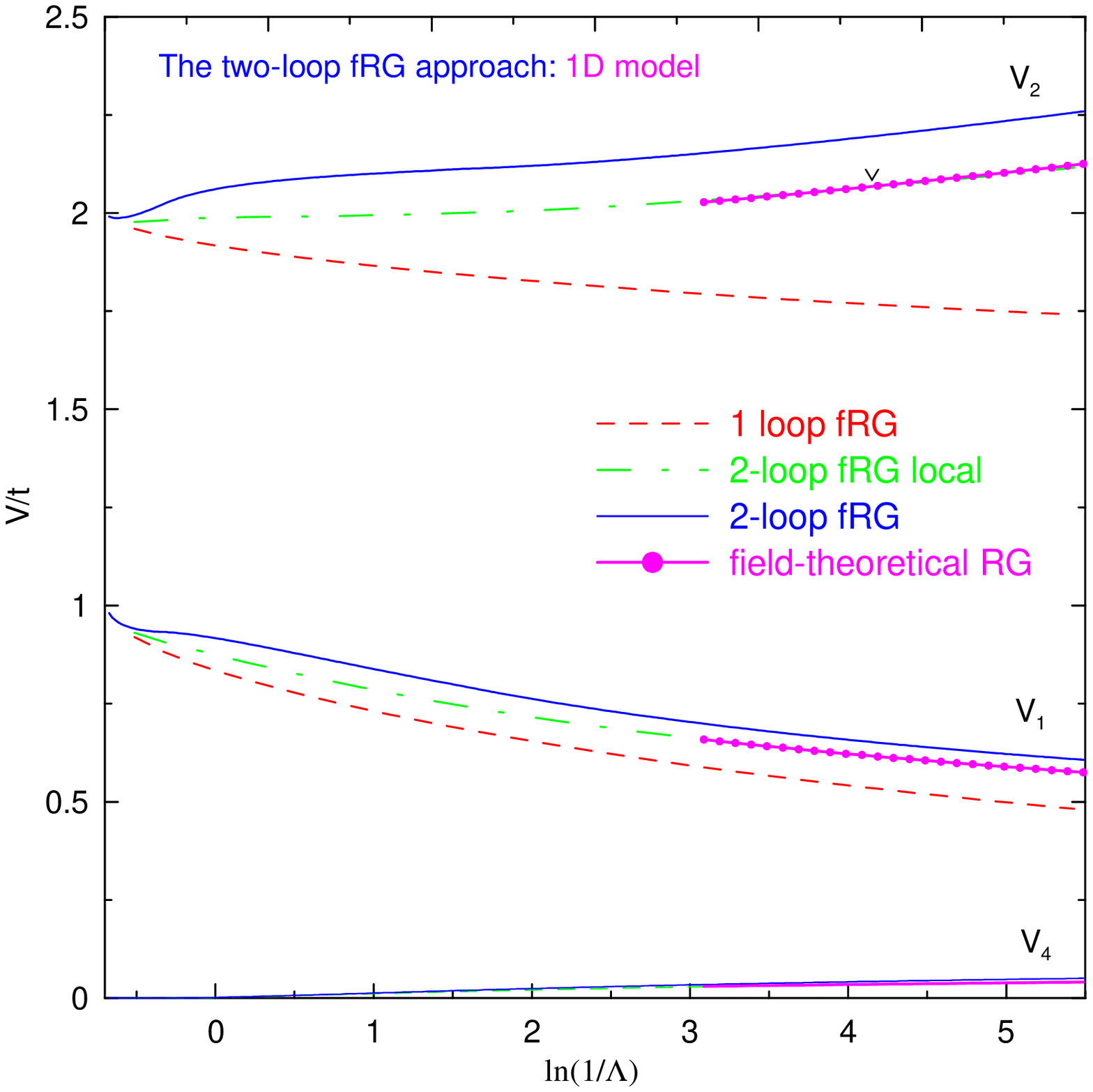} 
\includegraphics[width=8cm]{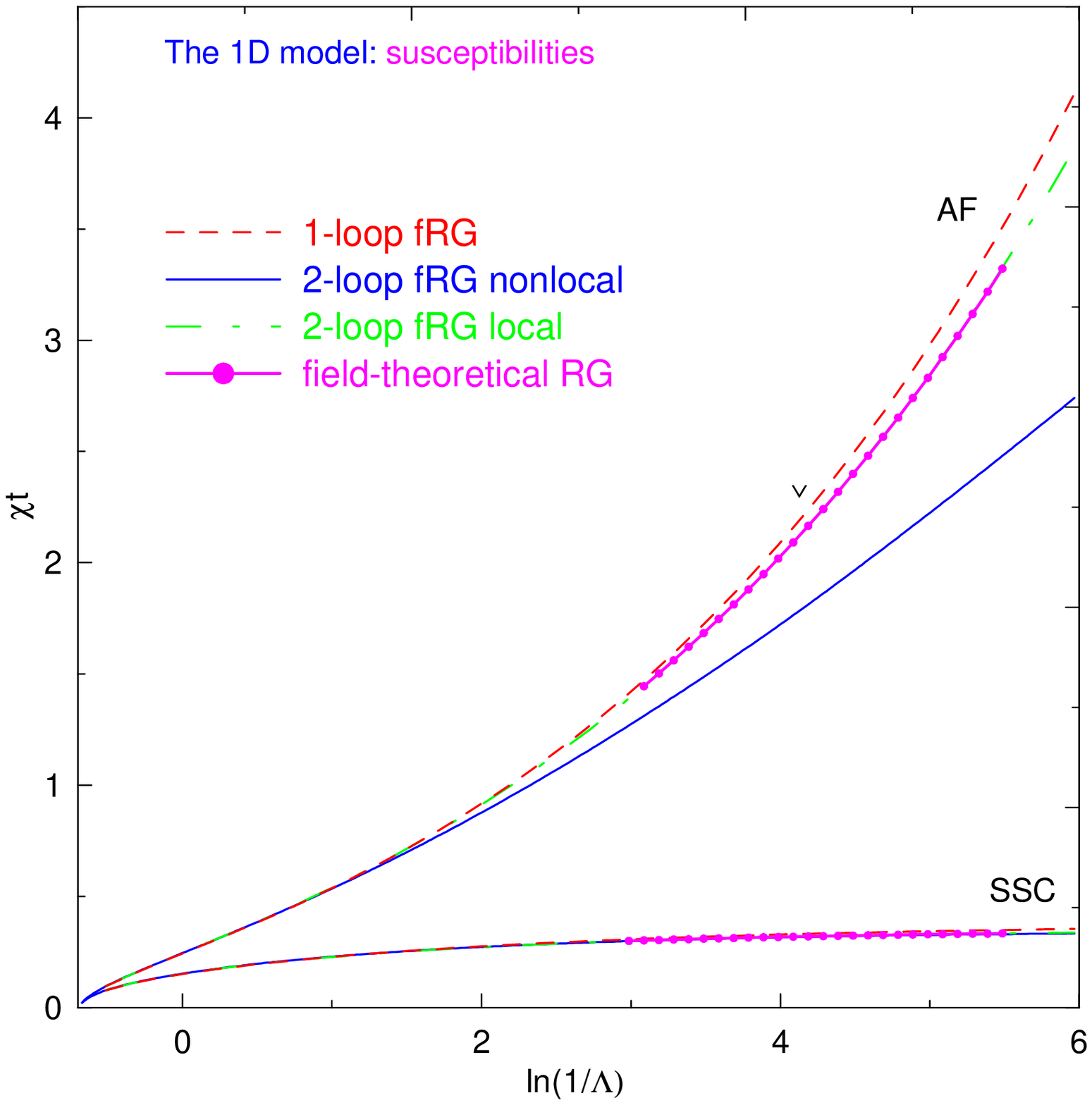} 
\caption{ (color online) The flow of the vertices (a), the antiferromagnetic
(AF) and the singlet superconducting (SSC) susceptibilities (b) of the 1D
Hubbard model within the non-local (solid lines) and local (dot-dashed
lines) two-loop approaches and the one-loop approach (dahed lines). The
results of the solution of the field-theoretical two-loop equations ( 
\protect\ref{SolyomEq}) with initial vertices, obtained in the fRG approach
at the scale $\Lambda =e^{-4}t$ (marked by arrow) are shown by bold lines
with circles.}
\end{figure}

\begin{figure}[tbp]
\includegraphics[width=8cm]{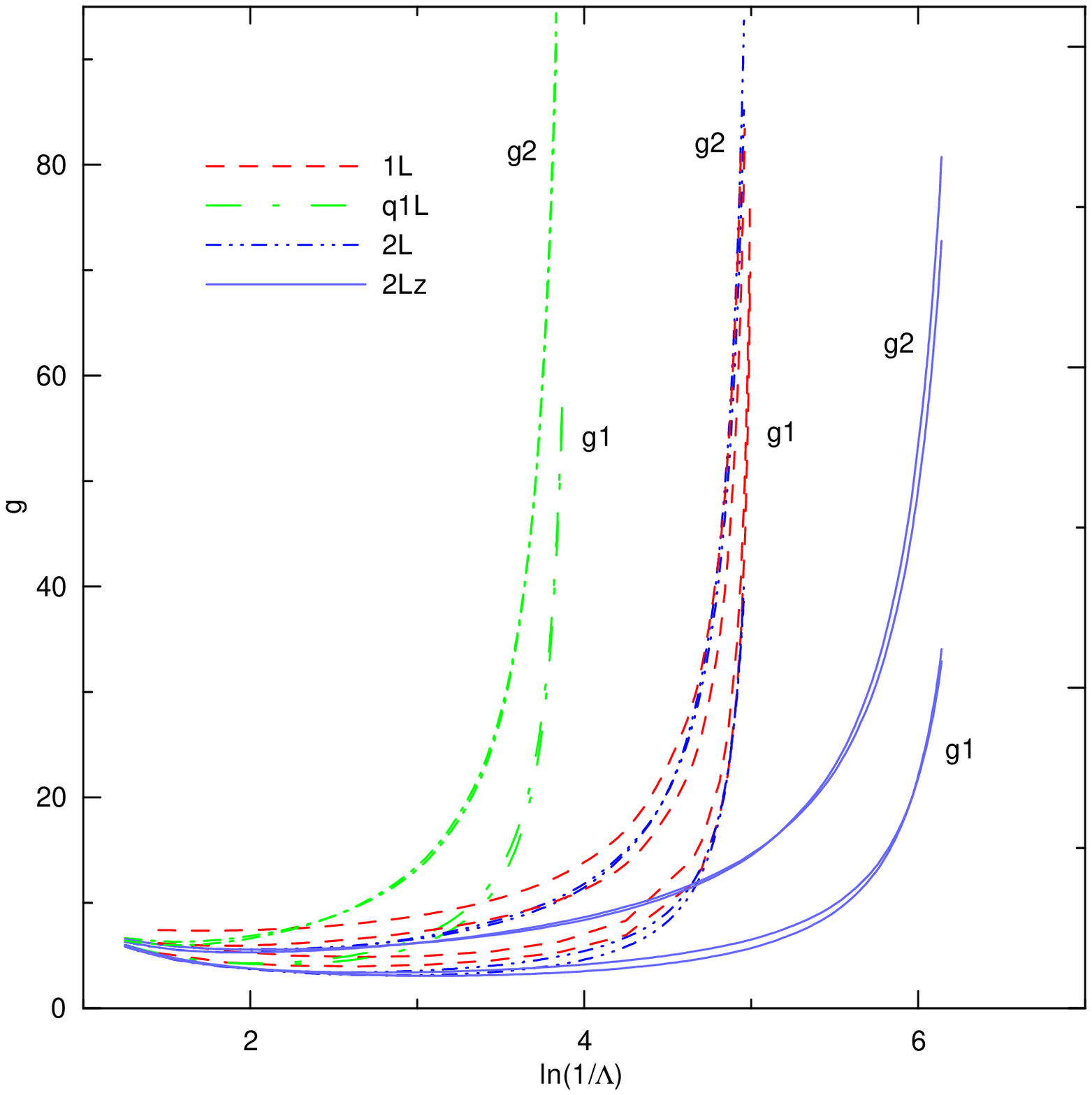} 
\caption{ (color online) The flow of the vertices $g_{1}=V(1,12,12)$ and $%
g_{2}=V(1,12,12)$ of the 2D Hubbard model with the flat Fermi surface in the
24-patch one-loop approach with vertex projection (1L, dashed lines), the
one-loop approach with partly corrected errors of the vertex projection
(q1L, dot-dashed lines) and the two-loop approach with account of Z-factors
(2Lz, solid lines) and without Z-factors (2L, dash-dot-dot lines). The first
and twelth patches being close to the centers of the opposite FS sides.}
\end{figure}

\begin{figure}[tbp]
\includegraphics[width=8cm]{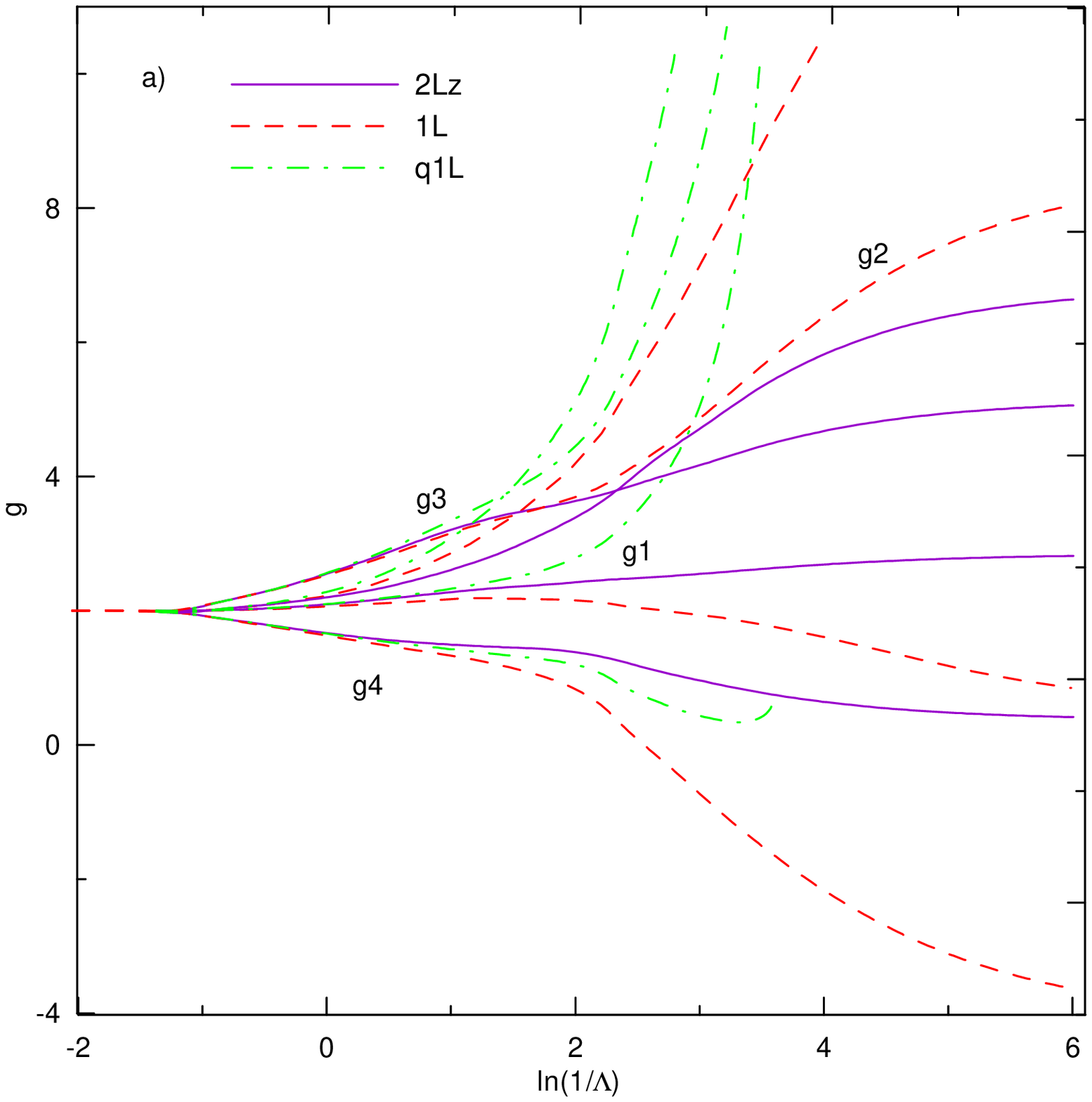} 
\includegraphics[width=8cm]{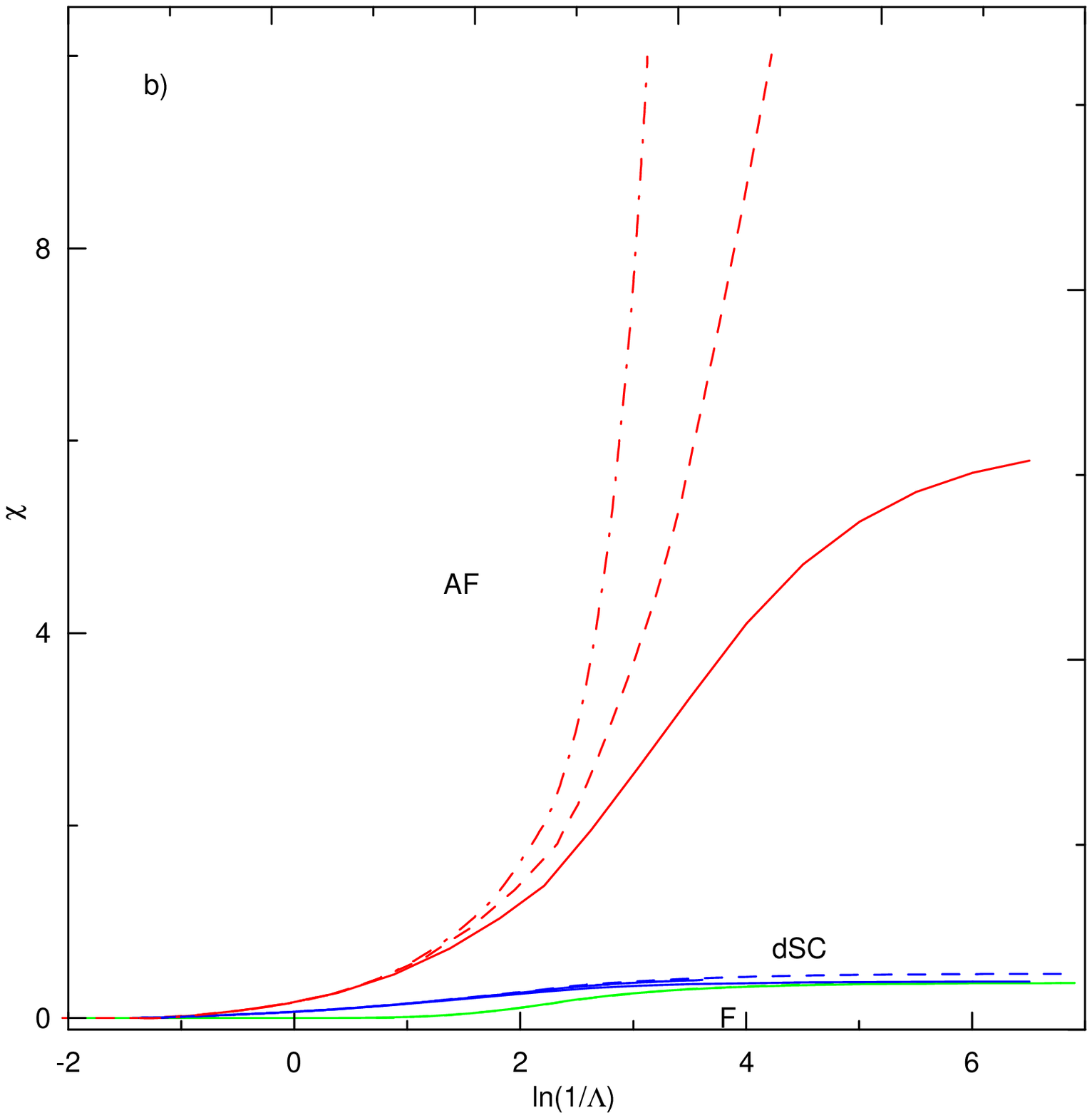} 
\includegraphics[width=8cm]{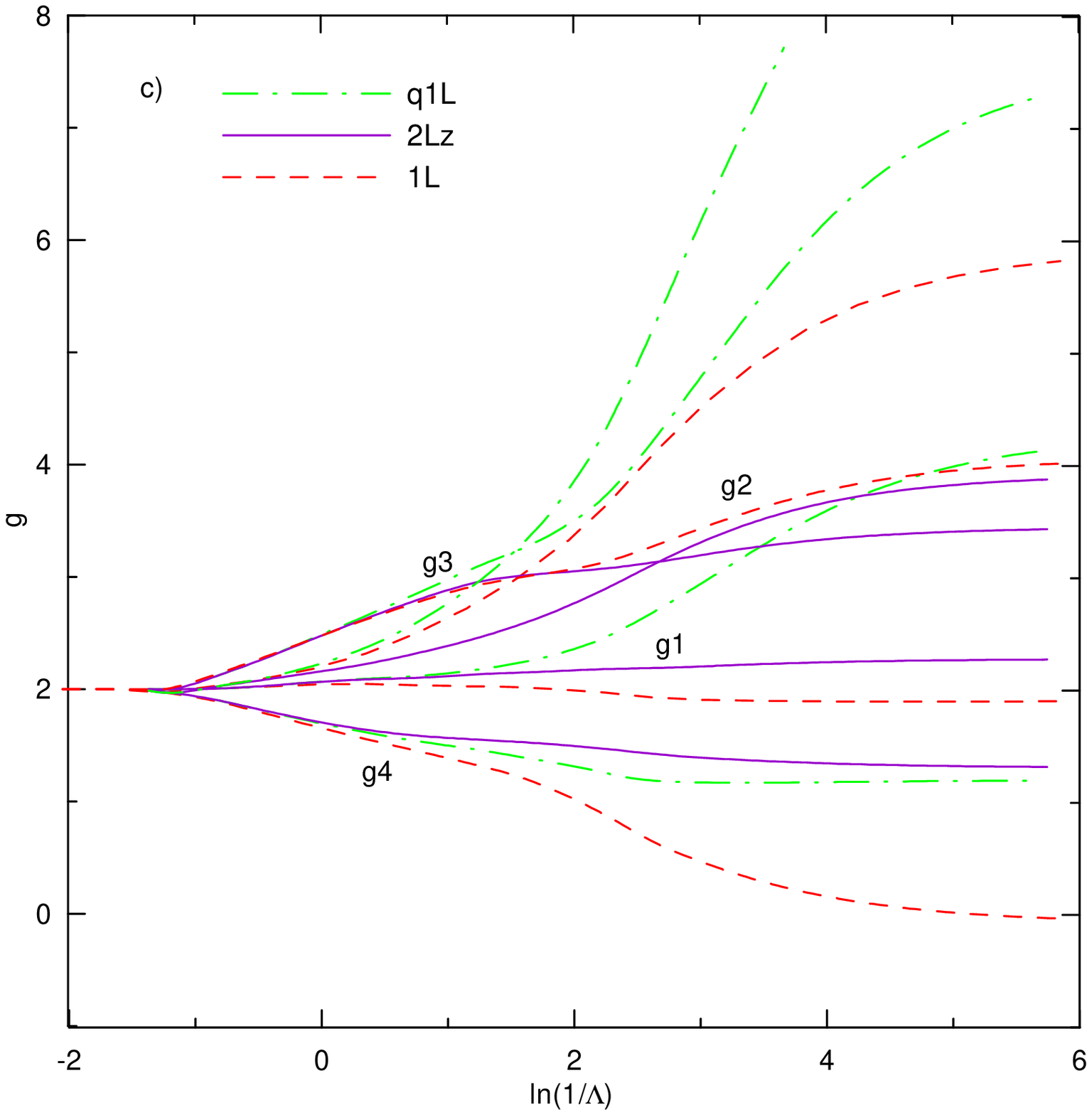} 
\includegraphics[width=8cm]{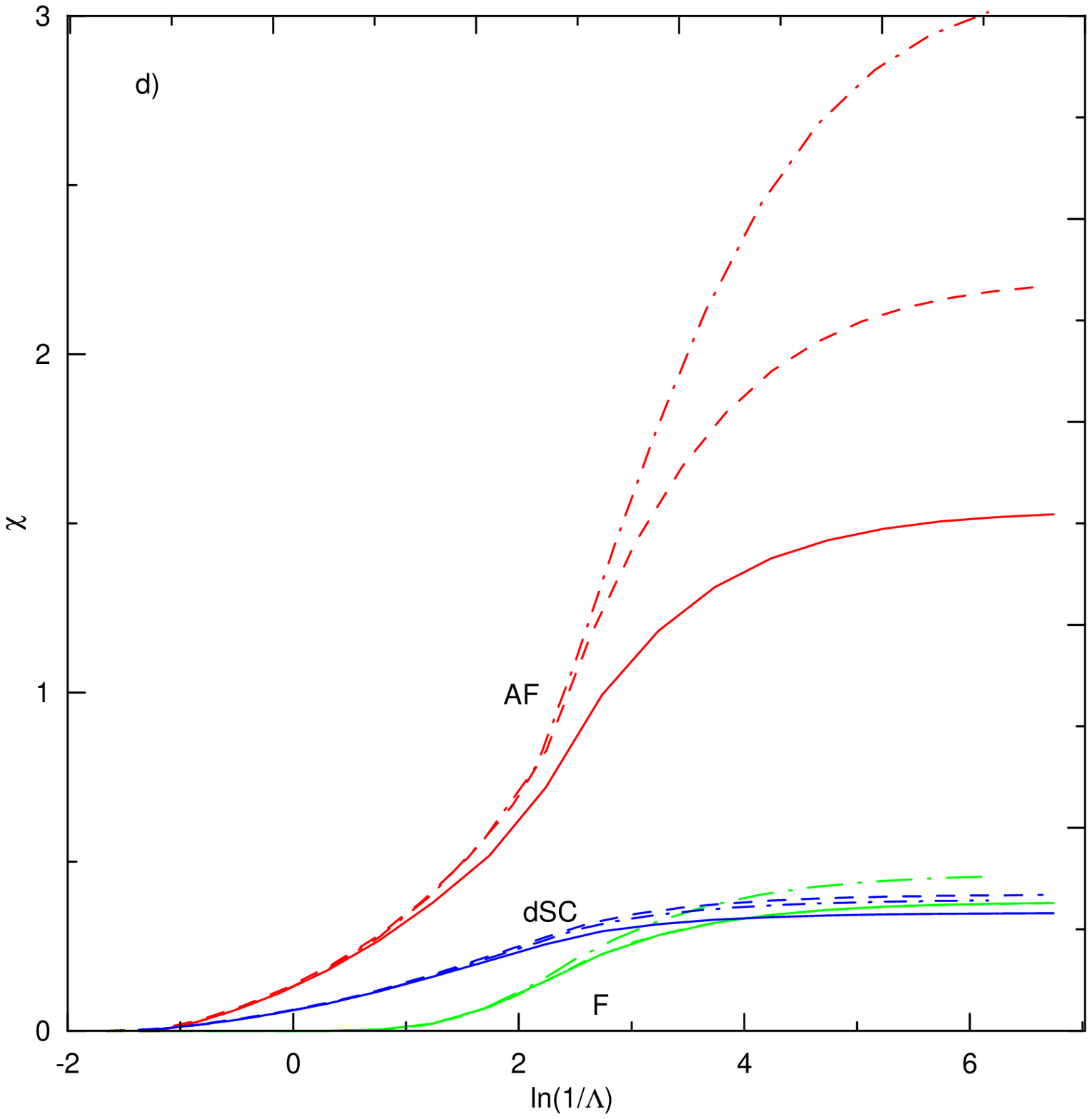} 
\caption{ (color online) The flow of the vertices $g_{1}=V(1,7,7)$, $%
g_{2}=V(1,7,1),$ $g_{3}=V(1,1,7),$ and $g_{4}=V(1,1,1)$ (a,c), the
antiferromagnetic (AF), d-wave superconducting (dSC) and the ferromagnetic
(F) susceptibilities (b,d) of the 2D $t$-$t^{\prime }$ Hubbard model with $%
U=2t,$ $t^{\prime }/t=0.1,$ $T=0.1t$ and $\protect\mu =0.1t$ (a,b), $\protect%
\mu =-0.1t$(c,d) in the 24-patch one- and two-loop fRG approaches. The first
and seventh FS patch correspond to points, closest to two different van Hove
singularities. The solid lines in b) and d) correspond to the two-loop
approach; the dashed lines - to the one-loop approach with projected
vertices, the dot-dashed lines - to the one-loop approach with partly
corrected errors of the vertex projections, other notations are the same as
in Fig. 5.}
\end{figure}

\begin{figure}[tbp]
\includegraphics[width=8cm]{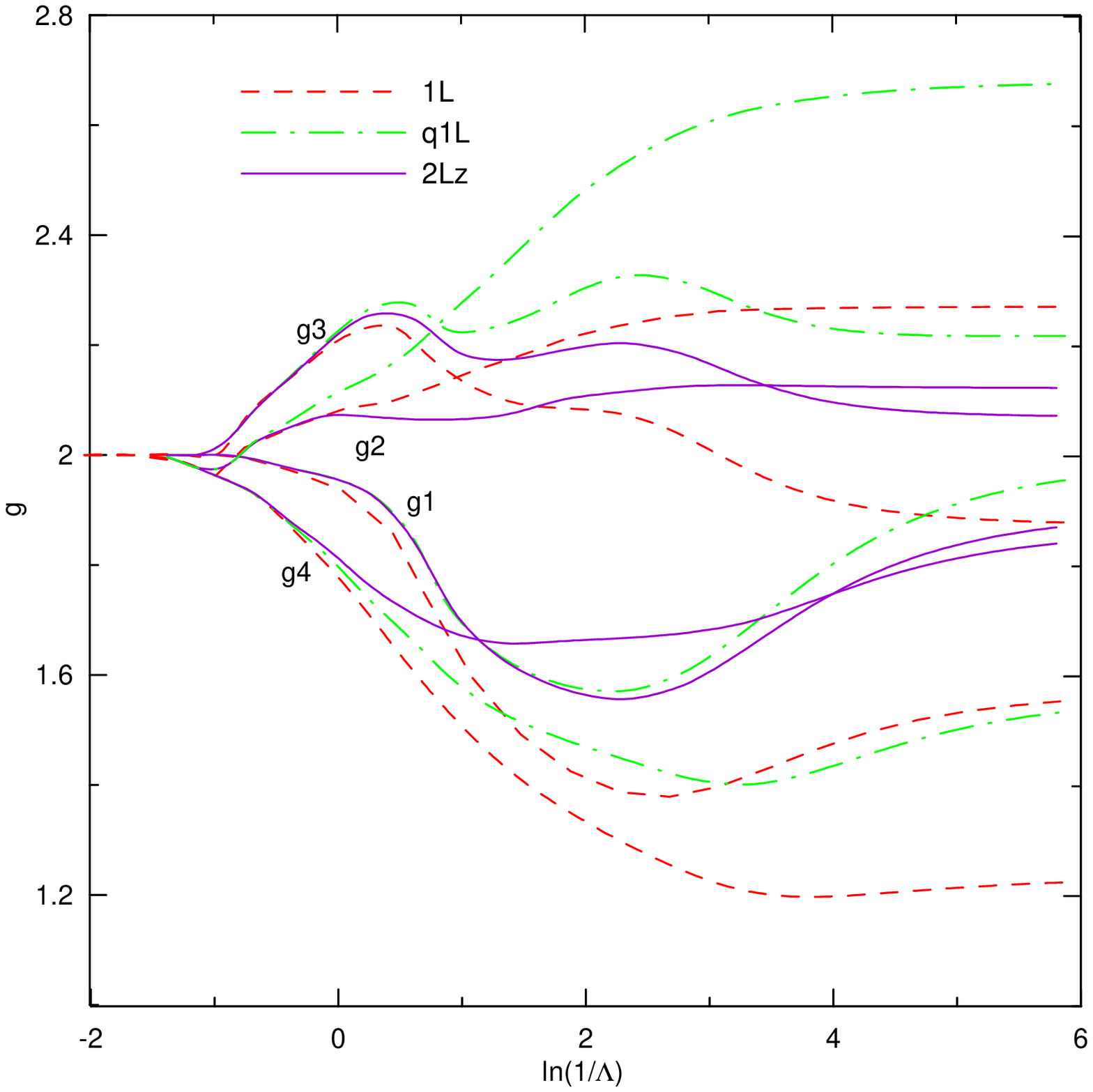} 
\includegraphics[width=8cm]{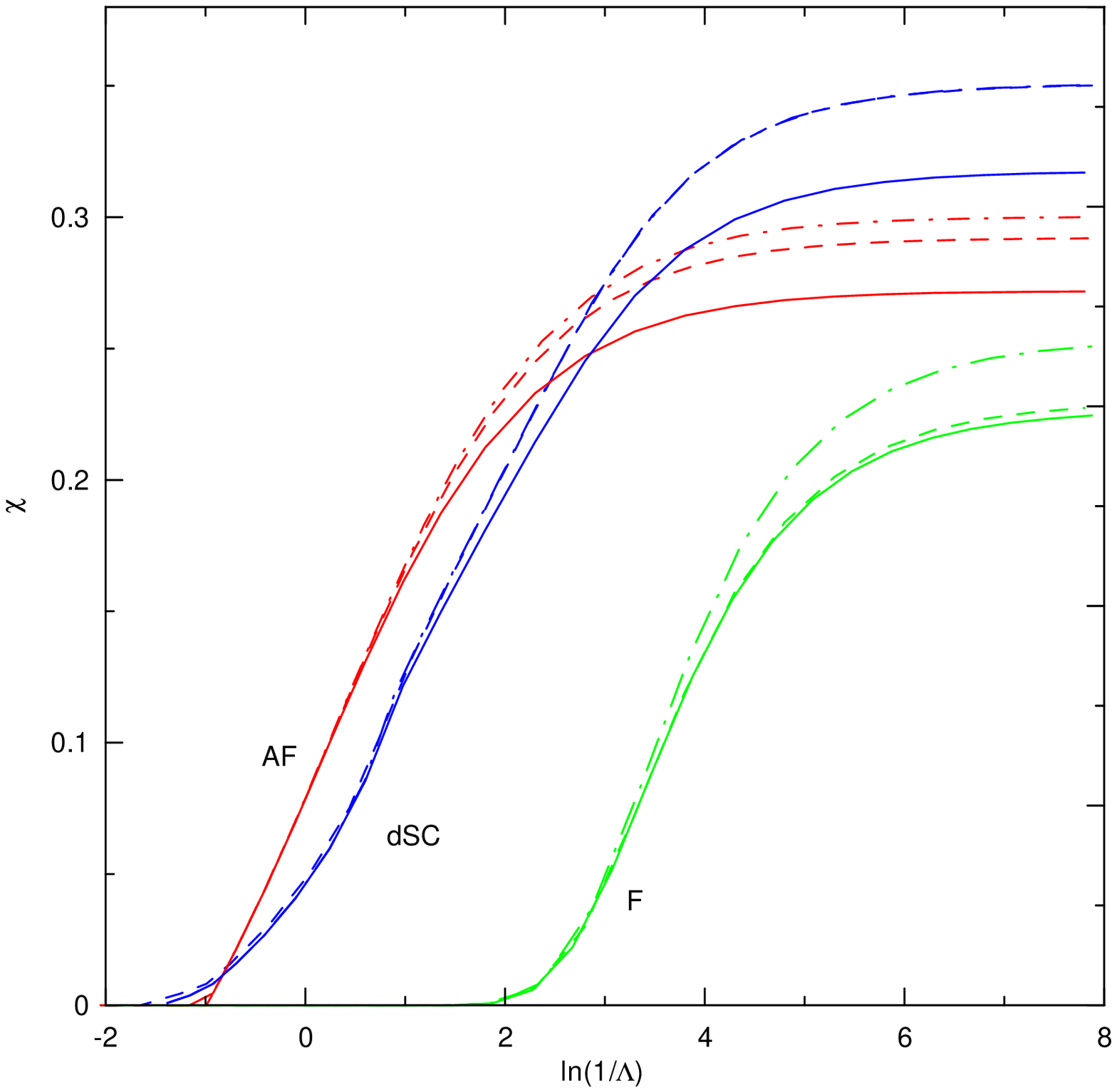} 
\caption{ (color online) The flow of the vertices (a) and the
susceptibilities (b) of the 2D $t$-$t^{\prime }$ Hubbard model, $U=2t,$ $%
t^{\prime }/t=0.1,$ $\protect\mu =-0.5t$, and $T=0.025t$ in the 24-patch
one- and two-loop fRG approaches, the notations are the same as in Fig. 6. }
\end{figure}

\begin{figure}[tbp]
\includegraphics[width=8cm]{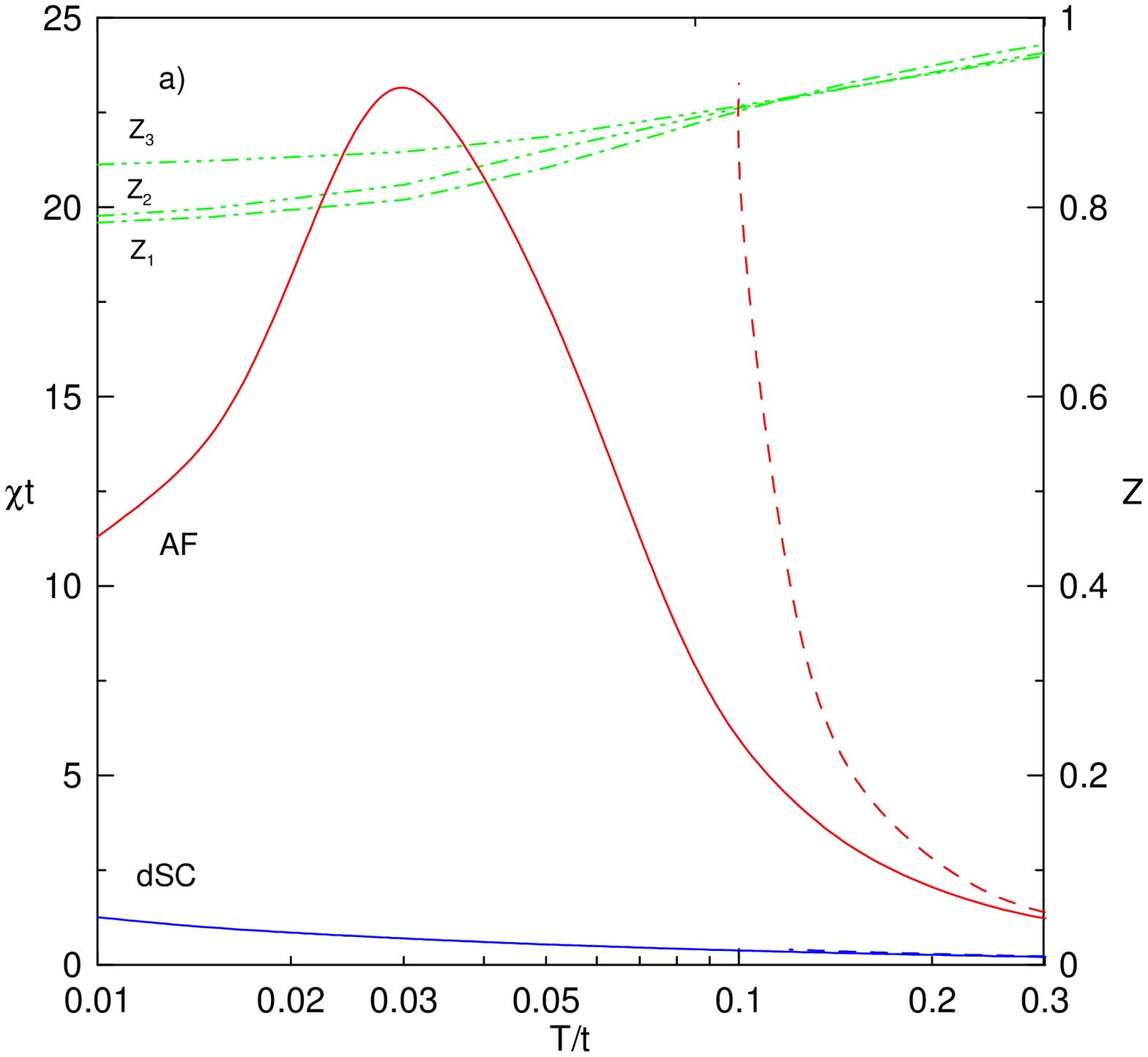} 
\includegraphics[width=8cm]{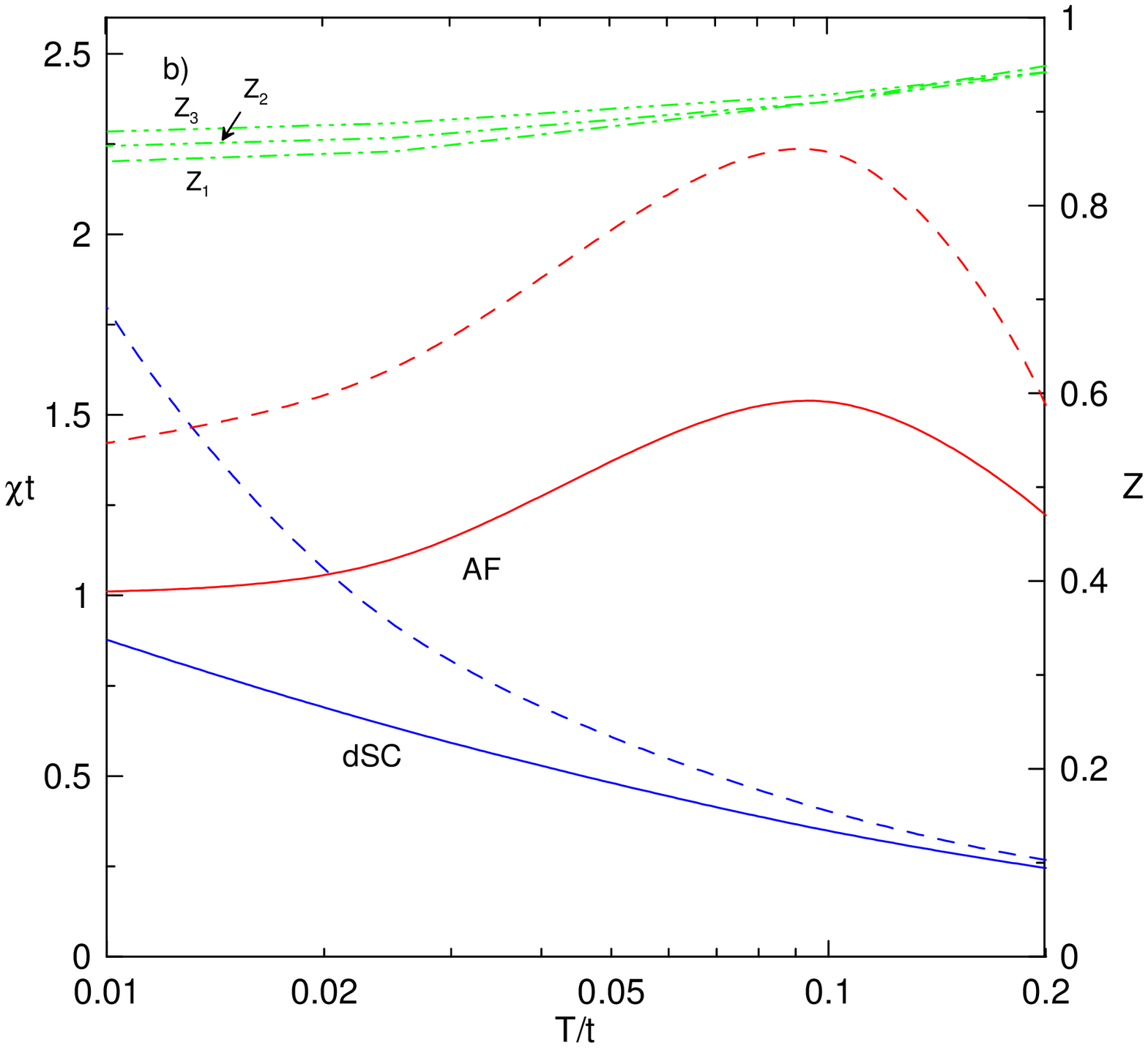} 
\caption{ (color online) The temperature dependence of the antiferromagnetic
and d-wave superconducting susceptibilities in one- (dashed lines) and
two-loop (solid lines) functional renormalization group approaches to the 2D 
$t$-$t^{\prime }$ Hubbard model with $U=2t$, $t^{\prime }/t=0.1$, $\protect%
\mu =0.1t$ (a) and $\protect\mu =-0.1t$ (b). The temperature dependence of
the Z-factors (right axis) is shown by dot-dashed lines; the number of dots
corresponds to patch number.}
\end{figure}

\begin{figure}[tbp]
\includegraphics[width=14cm]{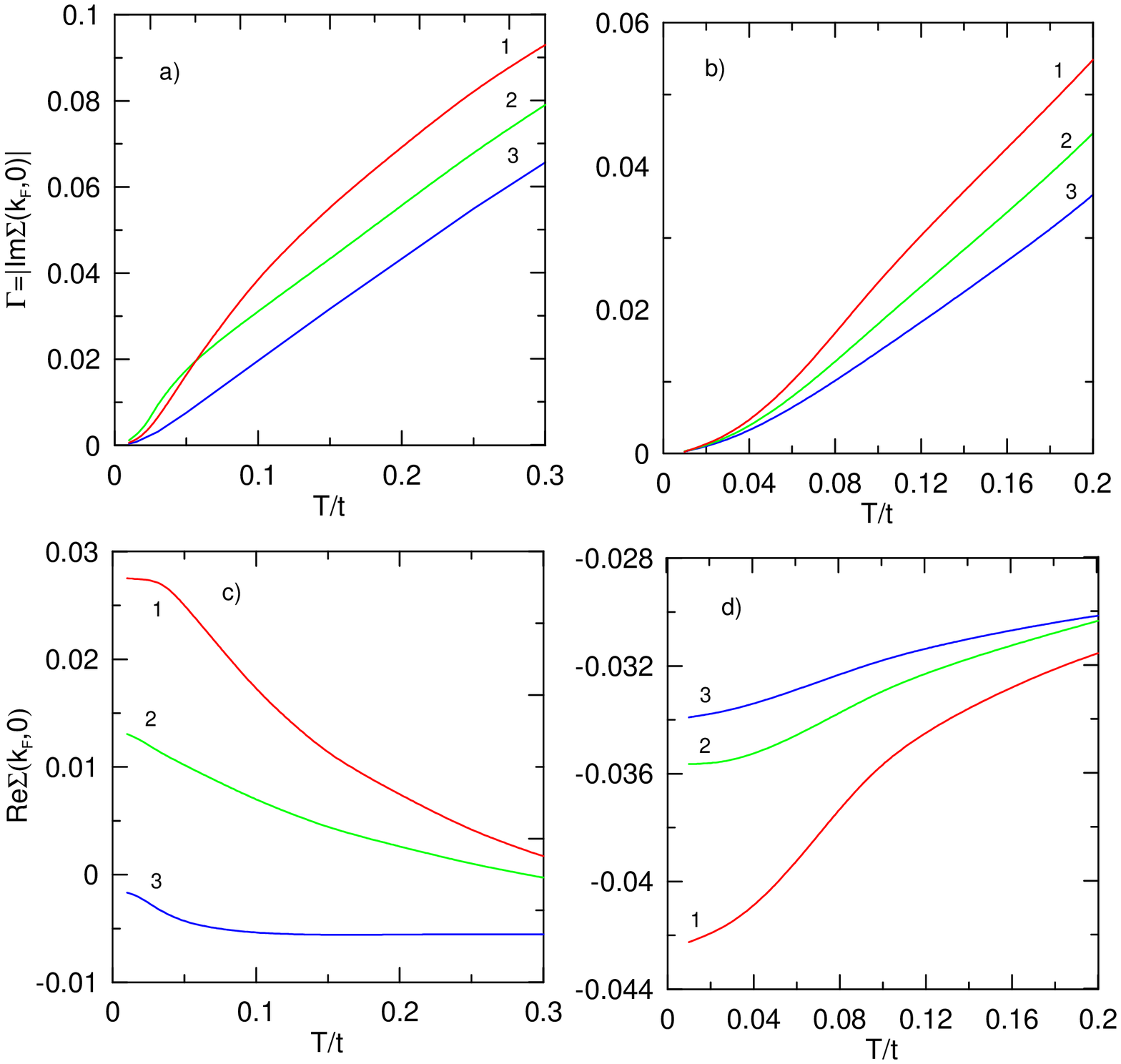} 
\caption{ (color online) The temperature dependence of the qp damping $%
\Gamma _{\mathbf{k}_{F}}=-\mathrm{Im}\Sigma (\mathbf{k}_{F},0),$ (a,b) and
the FS shift $\mathrm{Re}\Sigma (\mathbf{k}_{F},0)/t$ (c,d) at different FS
patches ($1,2,3$) obtained in the 24-patch two-loop fRG approach at $U=2t$
and $t^{\prime }/t=0.1,$ $\protect\mu =0.1t$ (a,c) and $\protect\mu =-0.1t$
(b,d). The patch 1 is the closest to the ($\protect\pi ,0$) point.}
\end{figure}

\end{document}